\documentclass[aps,prl,preprint,tightenlines,superscriptaddress,showpacs,byrevte
x]{revtex4}

\usepackage{graphicx} 
\usepackage{dcolumn}  

\graphicspath{{ps}}

\draft 

\begin{document}

\preprint{\vbox{ \hbox{   }
                 \hbox{BELLE-CONF-0519}
                 \hbox{LP2005-154}
                 \hbox{EPS05-491}
}}

\title{ \quad\\[0.5cm] 
Measurement of $D^0\rightarrow\pi l\nu (K l\nu)$ and their form
  factors}

\affiliation{Aomori University, Aomori}
\affiliation{Budker Institute of Nuclear Physics, Novosibirsk}
\affiliation{Chiba University, Chiba}
\affiliation{Chonnam National University, Kwangju}
\affiliation{University of Cincinnati, Cincinnati, Ohio 45221}
\affiliation{University of Frankfurt, Frankfurt}
\affiliation{Gyeongsang National University, Chinju}
\affiliation{University of Hawaii, Honolulu, Hawaii 96822}
\affiliation{High Energy Accelerator Research Organization (KEK), Tsukuba}
\affiliation{Hiroshima Institute of Technology, Hiroshima}
\affiliation{Institute of High Energy Physics, Chinese Academy of Sciences, Beijing}
\affiliation{Institute of High Energy Physics, Vienna}
\affiliation{Institute for Theoretical and Experimental Physics, Moscow}
\affiliation{J. Stefan Institute, Ljubljana}
\affiliation{Kanagawa University, Yokohama}
\affiliation{Korea University, Seoul}
\affiliation{Kyoto University, Kyoto}
\affiliation{Kyungpook National University, Taegu}
\affiliation{Swiss Federal Institute of Technology of Lausanne, EPFL, Lausanne}
\affiliation{University of Ljubljana, Ljubljana}
\affiliation{University of Maribor, Maribor}
\affiliation{University of Melbourne, Victoria}
\affiliation{Nagoya University, Nagoya}
\affiliation{Nara Women's University, Nara}
\affiliation{National Central University, Chung-li}
\affiliation{National Kaohsiung Normal University, Kaohsiung}
\affiliation{National United University, Miao Li}
\affiliation{Department of Physics, National Taiwan University, Taipei}
\affiliation{H. Niewodniczanski Institute of Nuclear Physics, Krakow}
\affiliation{Nippon Dental University, Niigata}
\affiliation{Niigata University, Niigata}
\affiliation{Nova Gorica Polytechnic, Nova Gorica}
\affiliation{Osaka City University, Osaka}
\affiliation{Osaka University, Osaka}
\affiliation{Panjab University, Chandigarh}
\affiliation{Peking University, Beijing}
\affiliation{Princeton University, Princeton, New Jersey 08544}
\affiliation{RIKEN BNL Research Center, Upton, New York 11973}
\affiliation{Saga University, Saga}
\affiliation{University of Science and Technology of China, Hefei}
\affiliation{Seoul National University, Seoul}
\affiliation{Shinshu University, Nagano}
\affiliation{Sungkyunkwan University, Suwon}
\affiliation{University of Sydney, Sydney NSW}
\affiliation{Tata Institute of Fundamental Research, Bombay}
\affiliation{Toho University, Funabashi}
\affiliation{Tohoku Gakuin University, Tagajo}
\affiliation{Tohoku University, Sendai}
\affiliation{Department of Physics, University of Tokyo, Tokyo}
\affiliation{Tokyo Institute of Technology, Tokyo}
\affiliation{Tokyo Metropolitan University, Tokyo}
\affiliation{Tokyo University of Agriculture and Technology, Tokyo}
\affiliation{Toyama National College of Maritime Technology, Toyama}
\affiliation{University of Tsukuba, Tsukuba}
\affiliation{Utkal University, Bhubaneswer}
\affiliation{Virginia Polytechnic Institute and State University, Blacksburg, Virginia 24061}
\affiliation{Yonsei University, Seoul}
  \author{K.~Abe}\affiliation{High Energy Accelerator Research Organization (KEK), Tsukuba} 
  \author{K.~Abe}\affiliation{Tohoku Gakuin University, Tagajo} 
  \author{I.~Adachi}\affiliation{High Energy Accelerator Research Organization (KEK), Tsukuba} 
  \author{H.~Aihara}\affiliation{Department of Physics, University of Tokyo, Tokyo} 
  \author{K.~Aoki}\affiliation{Nagoya University, Nagoya} 
  \author{K.~Arinstein}\affiliation{Budker Institute of Nuclear Physics, Novosibirsk} 
  \author{Y.~Asano}\affiliation{University of Tsukuba, Tsukuba} 
  \author{T.~Aso}\affiliation{Toyama National College of Maritime Technology, Toyama} 
  \author{V.~Aulchenko}\affiliation{Budker Institute of Nuclear Physics, Novosibirsk} 
  \author{T.~Aushev}\affiliation{Institute for Theoretical and Experimental Physics, Moscow} 
% ...
\collaboration{The Belle Collaboration}

\begin{abstract}

{Using a data sample of 282 fb$^{-1}$ collected by the Belle experiment at
the KEKB $e^+ e^-$ collider, we study $D^0\to \pi^- \ell^+ \nu$
and $D^0\to K^- \ell^+ \nu$ decays ($\ell$ = $\mu$,$e$) in $e^+e^-$
annihilation.
We identify $D^{*+}\to D^0\pi^+$ decays by using the mass of the system
recoiling against
a fully reconstructed tag-side $D^*$ or $D$ meson,
allowing for additional primary mesons from fragmentation.
Using a novel global reconstruction method that provides very good resolution in neutrino momentum and in $q^2 = (p_\ell+p_\nu)^2$, we reconstructed $D^{0} \rightarrow \pi^- \ell^+ \nu$ and $D^{0} \rightarrow K^- \ell^+ \nu$ decays.
From these events we measured the branching fraction ratios $BR(D^0 \to \pi e \nu)/BR(D^0 \to K e \nu) = 0.0809 \pm 0.0080 \pm 0.0032$ and $BR(D^0 \to \pi \mu \nu)/BR(D^0 \to K \mu \nu) = 0.0677 \pm 0.0078 \pm 0.0047$, and the semileptonic form factor ratio
$f_+({D^0 \to \pi^- \ell^+ \nu})^2/f_+({D^0 \to K^- \ell^+ \nu})^2\cdot|V_{cd}|^2/|V_{cs}|^2|_{q^2=0} = 0.041\pm 0.003\pm 0.004$ , where the errors are statistical and systematic, respectively. }
\end{abstract}

 
\maketitle
\tighten
{\renewcommand{\thefootnote}{\fnsymbol{footnote}}}
\setcounter{footnote}{0}

\section{Introduction}

Semileptonic decays of heavy-to-light mesons are of special 
interest since they are well suited to determine CKM matrix elements such as
$V_{ub}$, $V_{cd}$, $V_{cs}$.
However, some of these parameters --- in particular $V_{ub}$ ---are not yet measured with
satisfactory precision. 
The hadronic current in the semileptonic decays of $B$ and $D$ mesons is parametrized through 
form factors that depend on the invariant mass $q^2$ of the exchanged $W$ boson. In the past, the form factors 
for the two decays, $f_B(q^2)$ and $f_D(q^2)$, have been calculated in the quenched approximation of lattice QCD\cite{ref:1}.
Results from  
unquenched calculations have only recently become available \cite{ref:unquenched,ref:okamotopriv,ref:unquenched2}.
Imprecise knowledge of form factors is the main source of uncertainty in the extraction of the CKM matrix 
elements in semileptonic decays \cite{ref:PDG}.
When $|V_{ub}|$ is measured via $B^0\rightarrow\pi^- \ell^+\nu$, one uses  
the form factor $f_B(q^2)$; similarly, for $|V_{cd}|$ via $D^0\rightarrow\pi^- \ell^+\nu$, the form factor $f_D(q^2)$ is used.  
In the ratio of these form 
factors some of the theoretical uncertainties cancel \cite{ref:bdcancel}. Accurate 
information on the form factor $f_D(q^2)$ from the decay $D^0 \rightarrow \pi^- \ell^+\nu$ can be related to the measurement 
of $f_B(q^2)$ and hence also $|V_{ub}|$ 
and can test the validity of an effective pole ansatz \cite{ref:2} 
or the ISGW2 model~\cite{ref:3}.
The semileptonic form factor has also been investigated 
 by CLEO  ~\cite{ref:cleoff}, FOCUS ~\cite{ref:focus}, BES ~\cite{HQL4}, and CLEO-c~\cite{ref:cleoc}.

In addition to the semileptonic decay  $D^0\rightarrow\pi^- \ell^+\nu$ we also investigate 
the channel $D^0\rightarrow K^- \ell^+\nu$, 
which has much less background  and provides about 10 times higher statistics.
From a study of the two semileptonic channels we can evaluate 
the relative branching fractions of these decay modes. 

\section{Data and Monte Carlo Sets} 
The data collected by the Belle detector at the 
center of mass (CM) energy of 10.58 GeV ($\Upsilon(4S)$) and 60 MeV below 
that, corresponding to a total integrated luminosity of 282 
fb$^{-1}$, was used.
The Belle detector is a large-solid-angle magnetic spectrometer that 
consists of a multi-layer silicon
vertex detector (SVD), a 50-layer central drift chamber (CDC), an array 
of aerogel threshold
\v{C}erenkov counters (ACC), a barrel-like arrangement of time-of-flight 
scintillation counters
(TOF), and an electromagnetic calorimeter (ECL) comprised of CsI(Tl) 
crystals located
inside a superconducting solenoid coil that provides a $1.5$ T magnetic 
field. An iron flux-return
located outside of the coil is instrumented to detect $K^0_L$ mesons and 
to identify muons (KLM).
The detector is described in detail elsewhere \cite{Belle_det}. Two 
different inner detector
configurations were used. For the first sample corresponding to roughly half the statistics,  
a $2.0$ cm radius
beampipe and a 3-layer silicon vertex detector were used; for the second sample, 
a $1.5$ cm radius beampipe, a 4-layer silicon detector and a small-cell 
inner drift chamber were
used \cite{Belle2}.

Simulated Monte Carlo (MC) events were used for checks of the 
reconstruction method and background determination. The MC includes $\Upsilon(4S) \rightarrow B\overline{B}$ and 
continuum ($q\bar{q}$, where $q = c$, $s$, $u$, $d$) events, generated using the QQ 
generator~\cite{ref:bellegen} and processed through a complete GEANT-based simulation \cite{ref:bellemc} of the Belle detector. We refer to this  as the generic MC sample.
We also use MC samples of signal events for optimisation of selection criteria and signal reconstruction efficiency determination. 

\section{Analysis Strategy and Event Selection}

\begin{figure}
  \begin{center}
    \includegraphics[width=6.0cm]{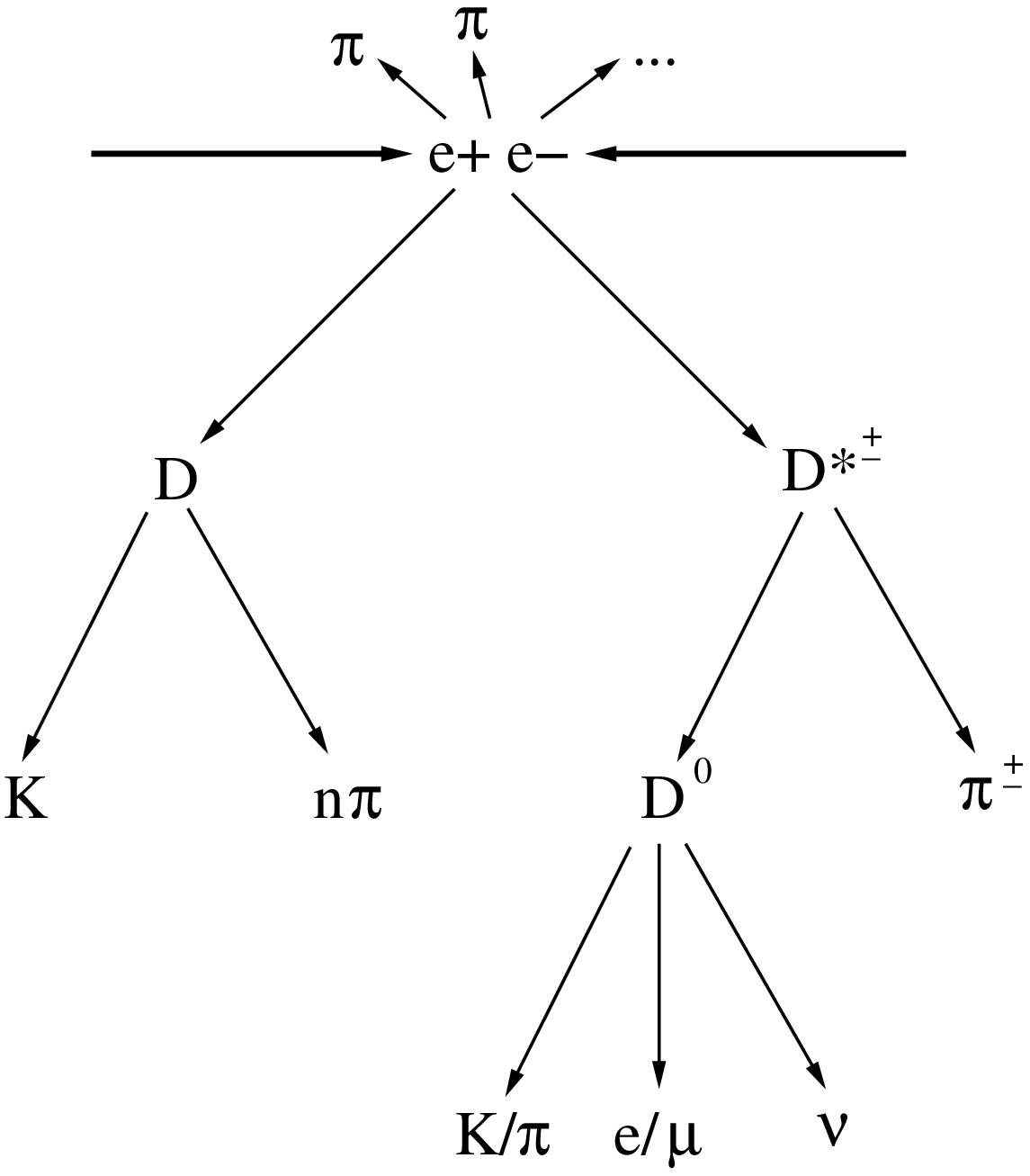}
    \includegraphics[width=6.0cm]{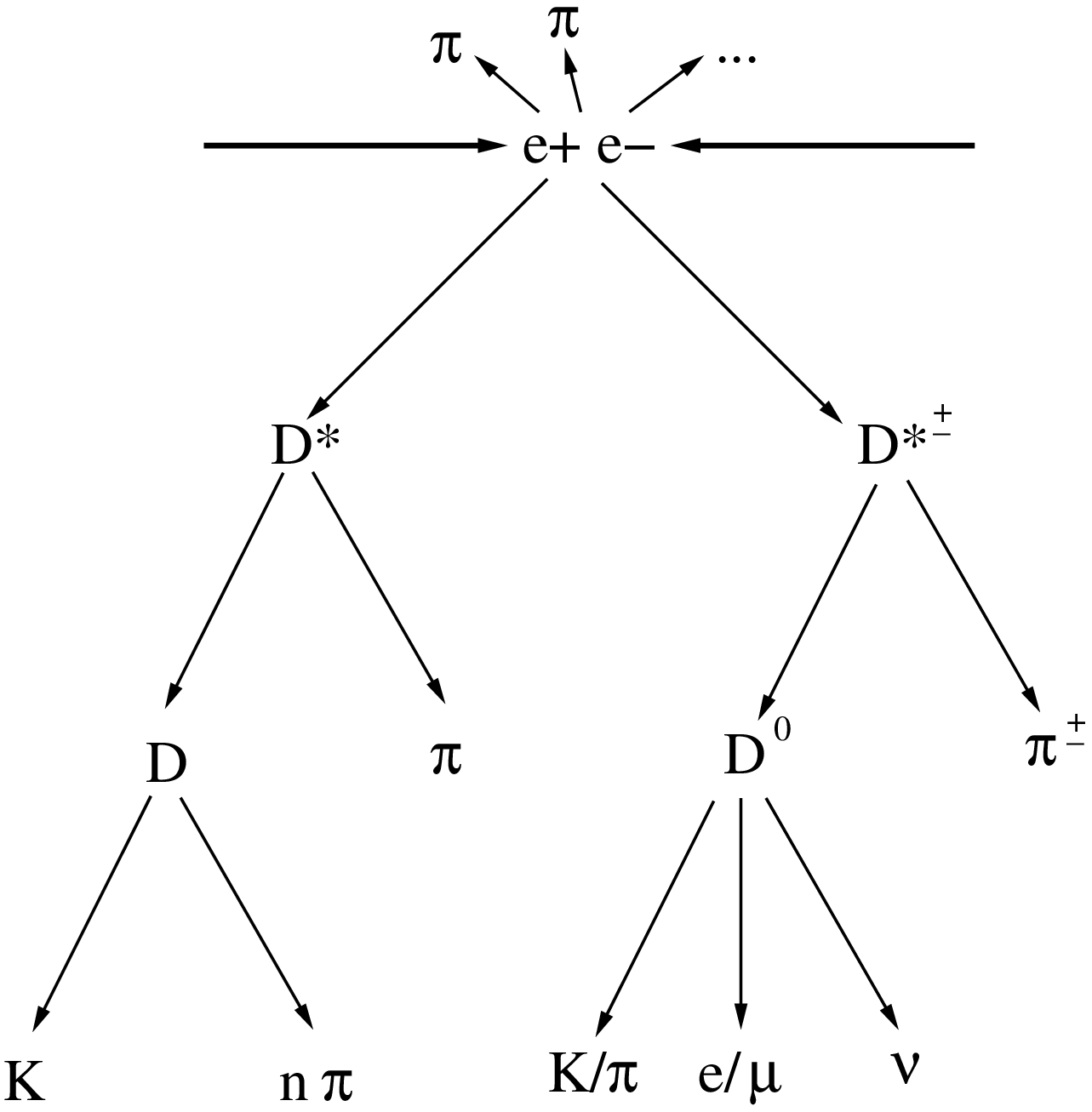}
  \end{center}
  \caption{Scheme of reconstructed event topology. Left: $DD^*$ events- D(tag)$D^{*\pm}$(signal). Right:  $D^*D^*$ events-$D^*$(tag)$D^{*\pm}$(signal).} 
\label{fig:recscheme}
\end{figure}

To measure semileptonic decays $D^0 \rightarrow h^-\ell^+\nu$
($h$ = $\pi$ or K, $\ell$ = $e$ or $\mu$), we consider event topologies 
of the type
$e^+e^- \rightarrow D^{(*)}D^*X$, where the $D^{(*)}$ system is referred to as the tag side, the $D^*$ system is referred to as the signal side, and $X$ 
denotes additional mesons including $\pi^0$, $\pi^\pm$ and $K^\pm$ (Fig. \ref{fig:recscheme}).
We use a global reconstruction technique that requires all the final
state particles in the event, with the exception of the neutrino from
the signal $D$ semileptonic decay, to be detected. 
In this paper, the reconstruction of charge conjugate modes is implied throughout.     
We found in agreement with \cite{ref:4} (which studied the case of no additional particles) that the $e^+e^- \rightarrow D\bar{D}X$ cross section is negligible, while for $DD^*$ and $D^*D^*$ the cross sections are so large that in spite of decreasing reconstruction efficiencies with increasing number of fragmentation particles, statistics can be significantly enhanced by our reconstruction method.
Instead of reconstructing decay chains  step-by-step by
applying intermediate selection criteria, all possible decay chains matching certain patterns 
(event topology and loose mass windows) are considered, leading to a possibly large number 
of intermediate combinations in a single event that is pruned at later steps in the reconstruction chain.
This strategy in many cases automatically solves the problem of ambiguities that would arise 
in a sequential reconstruction, and also allows looser selection criteria 
for individual particles, thus giving a higher 
reconstruction efficiency. 
Appropriate selection criteria, described 
below, are applied to select $D^0$ semileptonic decays.
The case of multiple signal candidates is treated by assigning each remaining candidate the same weight.

 Tracks are detected with the CDC and SVD. They are required to have  at least one associated hit in the SVD and an impact parameter with respect
 to the interaction point in the radial
 direction of less than $2$~cm and in the
 beam direction of less than $4$~cm.
 Tracks are also required to have momentum in the laboratory frame greater than $100$~MeV/$c$.
 A likelihood ratio for a given track to be a kaon or pion
 was obtained by utilising specific ionisation
 energy loss measurements made with the CDC, light yield
 measurements from the ACC, and time of flight information from the TOF.
 Lepton candidates were required to have momentum larger than 500
 MeV/$c$. For electron identification we use position, cluster energy,
 shower shape in the ECL, combined with track momentum and $dE/dx$ measurements in
 the CDC and hits in the ACC. For muon identification, we extrapolate
 the CDC track to the KLM and compare the measured range and
 transverse deviation in the KLM with the expected values. Photons are
 required to have energies in the laboratory frame greater than
 $50$~MeV. Neutral pion 
 candidates were reconstructed using $\gamma\gamma$ pairs with 
 invariant mass within $\pm 10$~MeV/$c^2$ of the nominal
 $\pi^0$ mass. These candidates are also subject to a mass constrained vertex fit, 
 assuming they arise from the interaction point, for which the
 confidence level is required to be greater than 0.1. 
   
 We reconstruct tag-side $D^\pm$ and $D^0$ candidates in the channels $D^\pm \rightarrow
 K^\pm n \pi$ and $D^0 \to K^\pm n \pi$ where the pions can be charged or neutral, and $n=1,2,3$. Depending on the channel the candidate
 mass window varies in the range $5-60$~MeV/$c^2$ around the nominal
 $D$ mass, corresponding to $\pm 5\sigma$. These candidates undergo a mass constrained vertex fit and
 we retain only those for which the fit is successful.  
 An attempt is then made to reconstruct $D^*$ candidates on the tag
 side via  $D^{*+} \to D^0\pi^+$, $D^{*+} \to D^+\pi^0$, $D^{*0}\rightarrow
 D^0 \pi^0$   , $D^{*0}\rightarrow D^0 \gamma$, and charge conjugate
 modes. The mass window around the nominal $D^*$ mass 
 is $5$ ($20$) MeV/$c^2$ for the non-radiative (radiative) mode.
 We require a successful mass constrained
 vertex fit of the $D^*$ candidate. 
 At this stage of the reconstruction we either have a fully
 reconstructed $D^{\pm 0*}$ or $D^{\pm 0}$ candidate on the tag side.

 Along with possible additional primary mesons, we search for signal side
 $D^{*\pm}$ in the recoil of the tag side:
the 4-momentum of the $D^{* \pm}$ candidate ($p_{D^{* \pm} {\rm cand}}$)
 is reconstructed
 using the 4-momenta of the beams $(p_{\rm beam})$,
 that of the tag-side $p_{D^{(*)}}$
 and of the primary meson system $p_{X}$ as
 $p_{D^{* \pm} {\rm cand}} = p_{\rm beam} - p_{D^{(*)}} - p_{X}$.
The accuracy of the $D^{* \pm}$ candidate momentum is improved by applying a
mass constrained vertex fit to the recoil particle. 
For candidates where the 
fit to $D^*$ has a confidence level probability larger than $0.1\%$, 
an analogous fit is repeated for a recoil $D^0$ from $D^{*\pm} \rightarrow D^0\pi^\pm_s$ decay. The 
charge of the slow pion $\pi_s$ is required to be the same as that of 
the kaon arising from the tag side $D$.
The small available phase space for $D^{*+} \rightarrow D^0\pi^+$ 
in addition to the mass constraint
leads to a very sharp peak in the $D^0$ candidate mass distribution (Fig. \ref{fig:gp_md0compd0}).
 Finally, neutrino candidates in the semileptonic $D^0$ decay are identified on the basis of the missing
 mass squared $m^2_{\mathrm{miss}}$, where the missing 4-momentum is calculated as 
\begin{equation}
	p_{\mathrm{miss}}=p_{D,\mathrm{signal-side}} - p_{K/\pi} - p_{e/\mu}.
\end{equation}
Neutrino candidates are retained if their CM energy is greater than $100$~MeV, if
the remaining photon candidates not used in the
reconstruction have an energy sum in the CM frame of less than
$700$~MeV and finally if the charges of the lepton and the slow pion are the same.

The signal region for $D^0$ semileptonic decays is defined
as $|m_\mathrm{miss}|^2<0.05\,\mathrm{GeV}^2/c^4$.
Our method yields a good resolution on the neutrino mass and 
the momentum transfer $q^2$ by applying a mass constrained vertex fit for neutrinos.
The latter is found to be $\sigma_{q^2} =
0.0145 \pm 0.0007 $ GeV$^2/c^2$ in MC signal events. 
 Although in the fit particles 
by default are assumed to originate from the interaction region, the influence of the $D^0$ decay distance 
on the reconstructed $q^2$ value has been corrected for by shifting the $D^0$ vertex in the direction of the fitted 
$D^0$ momentum by a distance corresponding to the mean lifetime of the meson. No significant change of the $q^2$ value or resolution 
has been observed. 

For future reference the multiplicity of an event is defined as the
total number of $\pi^\pm, \pi^0$ and $K^\pm$ mesons not assigned as decay
products of the tag-side $D$ or signal-side $D^*$. 

After the selection $75\%$ of remaining events have one candidate while a
negligible fraction have greater than three candidates.
Multiple candidates in events 
are mainly due to an interchange of particles within a  
particular decay chain, each is assigned equal weight within an event. A 
possible bias in the measurement of $q^2$ that may arise due to events where 
the lepton and meson are interchanged, a double mis-assignment, was 
checked with candidate $D^0 \to K^-\ell^+\nu$ events, and found to be negligible.  
 
\section{Background studies and Background subtraction}

\subsection{Non $D^0$-events and events with badly reconstructed $D^0$} \label{sec:nond0}
\begin{figure}
  \begin{center}
    \includegraphics[width=12.0cm]{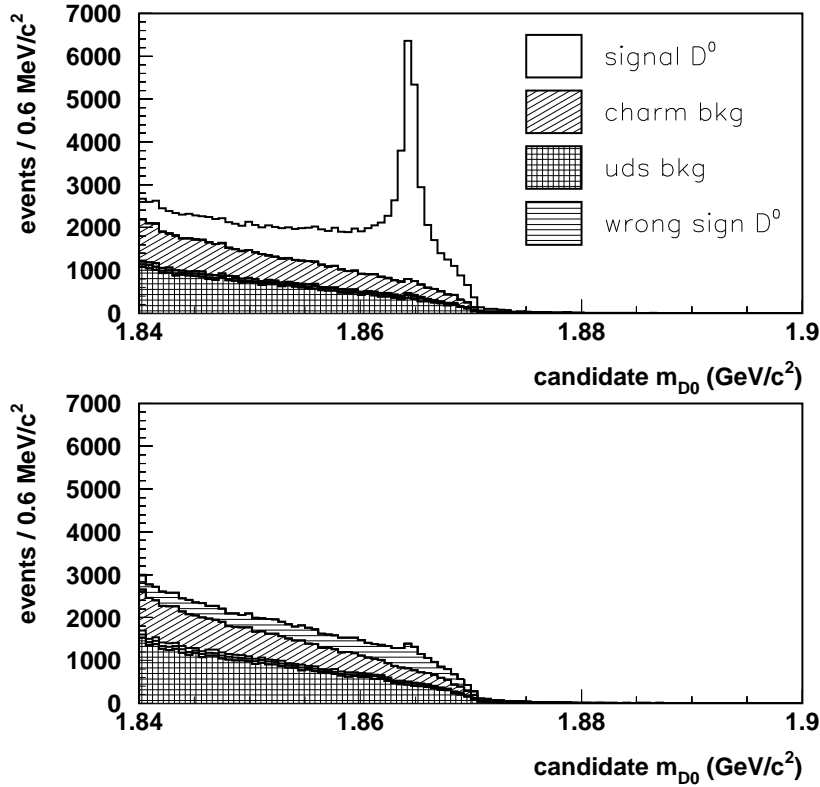}
  \end{center}
  \caption{Inclusive $D^0$ mass spectrum in generic MC events for right- (top) and wrong-sign (bottom)
    samples.The invariant mass shown is that prior to the $D^0$ mass-constrained fit.}
  \label{fig:gp_md0compd0}
\end{figure}

Fig. \ref{fig:gp_md0compd0} (top) shows the composition of the inclusive signal side $D^0$ invariant 
mass spectrum for generic MC, where the reconstruction of the neutrino candidate has
yet to be performed. The generic MC sample consists of events that may contain
$D^0$, namely $\Upsilon(4S)\rightarrow B\overline{B}$ (\emph{bottom}), $e^+e^-\rightarrow
c\bar{c}$ events (\emph{charm}), and a pure background sample of $e^+e^-\rightarrow q\bar{q}$
where $q=u,d,s$ (\emph{uds}). Signal events have been distinguished according
to MC generator information. 
Most of the background comes either from $uds$ or from a misreconstruction of the $D^0$. 
There is also a small contribution from the $B$-meson samples ($\approx 1.4\%$). 
Performing the full signal side reconstruction and applying all
signal selecting requirements as described above reduced the relative background level to
approximately $1\%$ ($10\%$) for $D^0 \to K^-\ell^+\nu$ ($D^0 \to \pi^-\ell^+\nu$), with the remaining background
dominated by charm events with an incorrectly reconstructed $D^0$. 
Several constrained fits that are performed in the reconstruction may make background
events look more like signal (peaking backgrounds). This possibility has been investigated in the generic MC sample and no peaking 
of the background has been observed.

The background is estimated by
taking the $D^0$ mass spectrum from a wrong-sign $D^0$ sample defined by 
opposite charges of the kaon from the tag-side $D$ decay  
and the signal-side slow pion.
The right- and wrong-sign $D^0$ mass spectra in generic MC are shown in 
Fig.\ref{fig:gp_md0compd0}.
Subtracting the wrong-sign $D^0$ mass distribution from the right-sign distribution results in almost complete cancellation of the background in the latter, implying that
the non-$D^0$ background is dominantly charge-uncorrelated \footnote{The slightly higher 
wrong-sign background in the $uds$-component can be understood as an effect of charge conservation:
the expected correlation of $n$ charges under the constraint of charge conservation, {\it i.e.,} a net charge of zero, but
otherwise uncorrelated, is $<Q_i Q_j> = -1/(n-1)$, which averages for the 
$uds$-sample to $-0.14 \pm 0.03$ compared to an observed correlation in the $uds$ $D^0$
background of $-0.16 \pm 0.01$.}. 
The bottom plot in Fig. \ref{fig:gp_md0compd0} shows that $D^0$ candidates are also found in the wrong sign sample, with a yield approximately $10\%$ of that in the right-sign sample.   

Since the charge-correlated non-$D^0$ background 
is very small, only the charge-uncorrelated part is considered, and a systematic error is assigned for 
the charge-correlated part, defined as the difference of background in right- and wrong-sign, equal to its relative contribution as observed in MC. 
To subtract the small component of real $D^0$'s in the wrong-sign sample, we use the difference of the right- and 
wrong-sign sample as model of the true $D^0$ signal shape, and fit this to the wrong-sign sample. For this purpose, 
the wrong-sign background is approximated as a second order polynomial in the signal region, $1.862-1.867$ GeV/$c^2$.
The fitted wrong-sign $D^0$ component is subtracted, and the remaining distribution is used to represent 
the shape of the non-$D^0$ background. 
It is normalized to the data 
in the sideband $1.84-1.85$ GeV/$c^2$ of the $D^0$ mass spectrum 
thus obtaining the amount of non-$D^0$ background in the signal region 
(which is defined as the region selected by the criterion on the confidence level of the $D^0$ 
mass-constrained vertex fit). 
This method is confirmed with
generic MC (using $D^0 \to \pi^-\ell^+\nu$ selection criteria):
 the measured background is $6.5 \pm 1.4$ events, compared 
to the true value of $6$.

Yields of the measured non-$D^0$ 
background
are given in Table \ref{tab:events}. The systematic error for this measurement consists of a statistical part 
(from background shape and sideband statistics), a part due to the subtraction of wrong-sign signal events, 
the effect of the small component of charge-correlated background measured with MC, and of possible bias of 
the background shape due to selection criteria, which has been estimated by comparing the measurement with 
the result using loose selection criteria, {\it i.e.,} only those selecting $D^0$ (without the requirement on the confidence level of the mass constrained fit).
Individual uncertainties are listed in Table \ref{tab:syst}.

\subsection{Background from semileptonic decays with a misidentified or undetected meson} \label{sec:semi}

The background from real semileptonic decays, with an incorrectly identified meson or additional 
mesons lost in reconstruction, is highly suppressed by the good neutrino mass resolution (see measurements below).
For the signal channel $D^0 \rightarrow \pi^-\ell^+\nu$, there are backgrounds from 
$D^0 \rightarrow K^-\ell^+\nu$, $D^0 \rightarrow \rho^- \ell^+ \nu$ (with $\rho^- \rightarrow \pi^-\pi^0$), 
$D^0 \rightarrow K^{*-} \ell^+ \nu$ (with $K^{*-} \rightarrow K^0\pi^-$), as well as non-resonant 
decays into the latter two final states.  
The most significant background is due to $D^0 \rightarrow K^- \ell^+ \nu$ since the branching 
fraction is approximately 10 times larger than the signal.

For the signal channel $D^0 \rightarrow K^- \ell^+ \nu$, it was verified using MC that 
backgrounds from $D^0 \to \pi^- \ell^+ \nu$
and $D^0 \to \rho^- \ell^+ \nu$ are
completely negligible due to the much smaller branching fractions.
The latter channel is suppressed even further by the selection criteria, due to the additional missing momentum taken by the $\pi^0$ from $\rho^- \rightarrow \pi^-\pi^0$.
The channel $D^0 \rightarrow K^{*-} \ell^+ \nu$ contributes a small amount via $K^{*-} \rightarrow K^-\pi^0$.

To measure the semileptonic background  in the $D^0 \to \pi^- \ell^+ \nu$ sample from $D^0\rightarrow K^-\ell^+\nu$, the distribution of simulated $D^0 \to K^-\ell^+\nu$ 
decays that pass the selection criteria was normalized to the signal peak of the $D^0 \to K^-\ell^+\nu$ data events,
then multiplied by the expected misidentification probability for kaons to be identified as pions. The misidentification rates were corrected for known differences with 
real data, measured with a dedicated $D^*$ sample in bins of 
meson momentum and 
polar angle. 
The remaining 
contribution of $D^0 \rightarrow K^{*-} \ell^+ \nu$ and $D^0 \rightarrow \rho^- \ell^+ \nu$ was obtained 
from MC and normalized in the region $m^2_\nu>0.3\ \mbox{GeV}^2/c^4$. 
For $D^0 \to K^-\ell^+\nu$, the only semileptonic background stems from $D^0 \to K^{-*}\ell^+\nu$, which was determined with normalized MC as described above.

The systematic error in the estimate of the semileptonic background includes uncertainties due to MC statistics, 
in the misidentification probability correction and in the ratio of $K^*$ and $\rho$ contributions.
Yields of the semileptonic background are shown in Table \ref{tab:events}; their 
contributions to the systematic error are listed in Table \ref{tab:syst}.

\subsection{Hadronic $D^0$-decays with fake leptons} \label{sec:had}

As the fake rate for muons is about an order of magnitude larger than that for electrons, 
the background from hadronic $D^0$-decays with fake muons is much more significant.   
To study and further categorize this kind of background, it is useful to look once more at wrong-sign data, in this case
the lepton has opposite charge to that of the slow pion. 
The large $D^0 \rightarrow K^- \pi^+$ channel can
contribute to the background for $D^0 \rightarrow K^- \ell^+ \nu$ (right-sign), and $D^0 \rightarrow \pi^- \ell^+ \nu$ 
(wrong-sign) \footnote{In the first case, the $\pi^+$ is misidentified as lepton; in the second case, 
the $K^-$ is misidentified as lepton.}.  The wrong mass
hypothesis allows some energy for a possible neutrino candidate, but with the
additional requirement $E_{\mathrm \nu} \ge$ 100 MeV this background is removed for both cases.
However, in the decay $D^0 \rightarrow K^- \pi^+ \pi^0$ the $\pi^0$ may have enough momentum to fake 
a neutrino that passes the $E_{\mathrm \nu}$ selection criterion. For $D^0 \rightarrow K^- \ell^+ \nu$ ( $D^0 \rightarrow \pi^- \ell^+ \nu$), 
this background contributes to the right-sign (wrong-sign) sample.
The background channel $D^0 \rightarrow \pi^- \pi^+ \pi^0 $ 
contributes equally to right and wrong sign samples.

\begin{figure}
  \begin{center}
    \includegraphics[width=8.0cm]{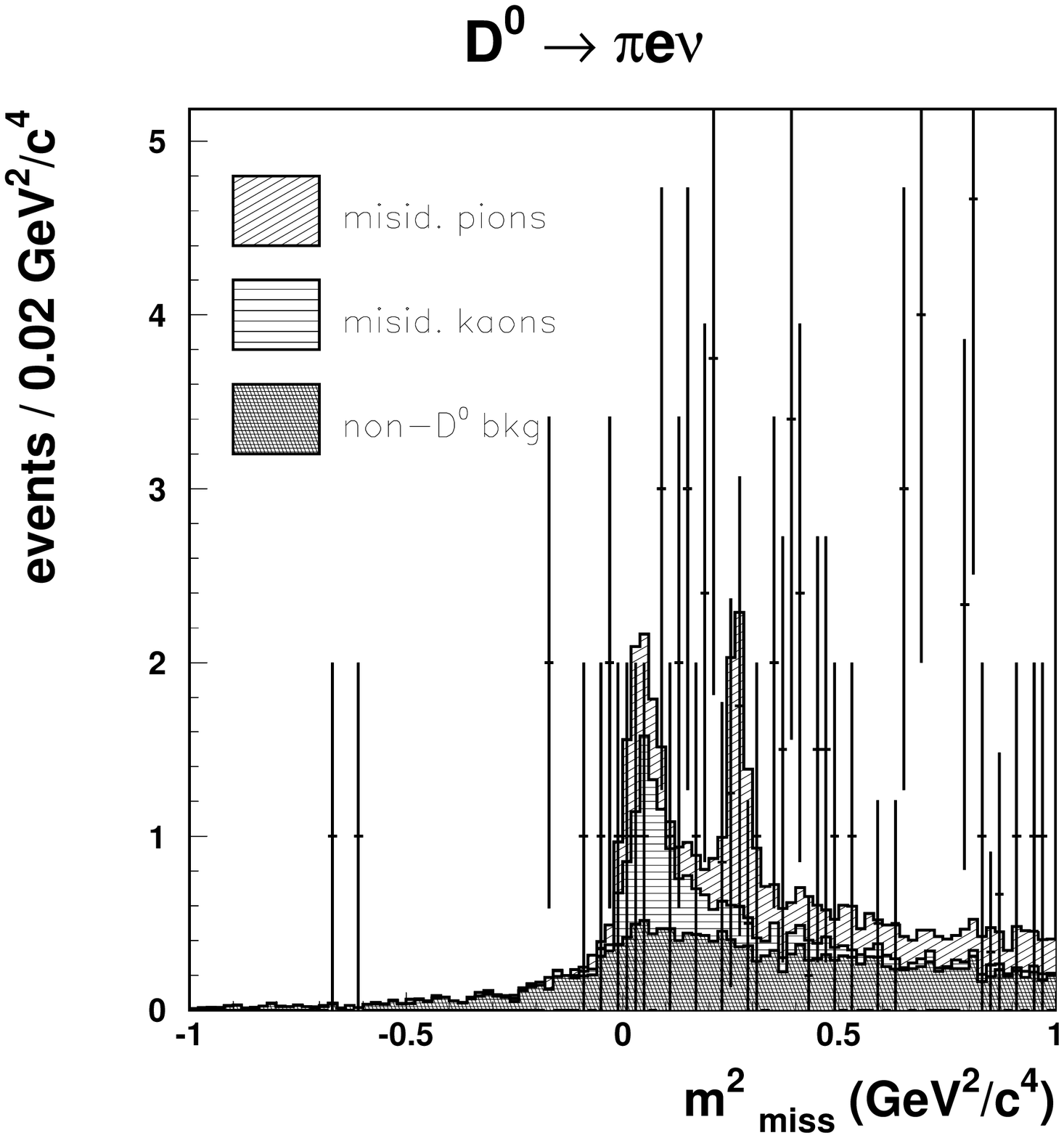}
    \includegraphics[width=8.0cm]{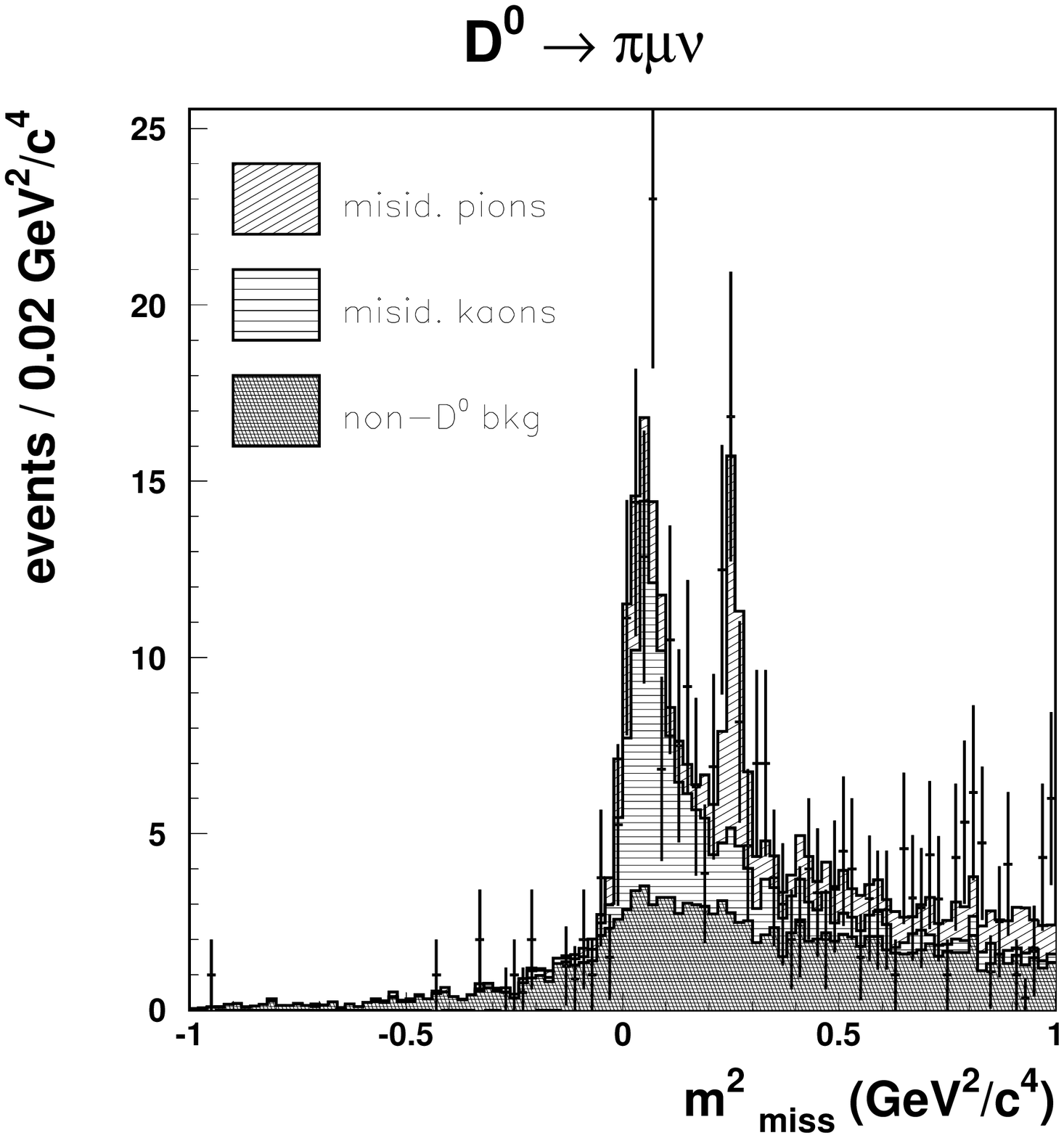}
  \end{center}
  \caption{Fitted hadronic background shapes for the pion channels in the wrong-sign data sample (defined by 
opposite charges of the slow pion and the lepton); the two prominent peaks correspond to background from $D^0 \to K^- \pi^+ \pi^0$  (lower $m_\mathrm{miss}^2$) and $D^0 \to K^0 \pi^+ \pi^-$ (higher $m_\mathrm{miss}^2$)}
\label{bkg2}
\end{figure}

To measure the shapes of these backgrounds, special samples are used wherein identified kaons
and pions are intentionally misidentified as leptons.
As these raw shapes also contain some amount of non-$D^0$ background, 
the same procedure as described in the previous section is used to 
measure and subtract this part. 
A linear combination of the shapes of intentionally misidentified kaons and 
pions is fitted to the measured $m_\nu^2$ distribution of the wrong-sign sample (see Fig. \ref{bkg2}). 
Assuming that the $K^\pm$ and $\pi^\pm$ misidentification rates do not depend on the 
charge of the slow pion, the same amount of background contributes to the wrong-sign 
and the right-sign samples.
Since the statistics for wrong-sign data samples in the $D^0 \to K^-\ell^+\nu$ channel 
is very small, we use the measured fake-rates in $D^0 \to \pi^-\ell^+\nu$ 
for the kaon modes as well. Numerical results can be found in Table \ref{tab:events}.

It has been checked using the generic MC sample that the shape of the
background is relatively insensitive to the differences in the kinematics
of correctly identified and mis-identified mesons; the possible small differences have been
included in the systematic error (see Table \ref{tab:syst}). The statistical error in the fits to the wrong-sign 
sample is included in the systematic uncertainties as well. 

\begin{figure}
  \begin{center}
    \includegraphics[width=8.0cm]{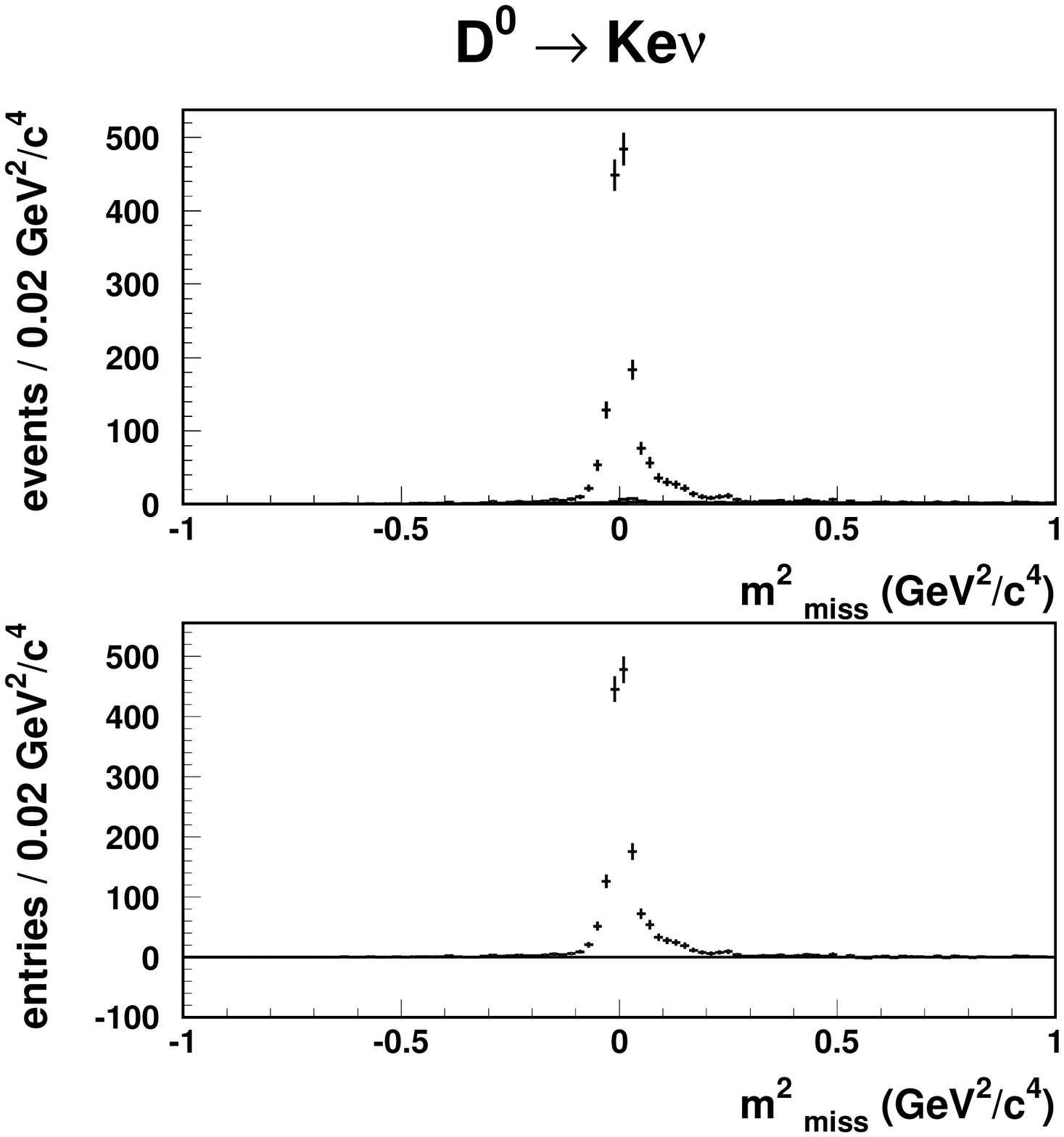}
    \includegraphics[width=8.0cm]{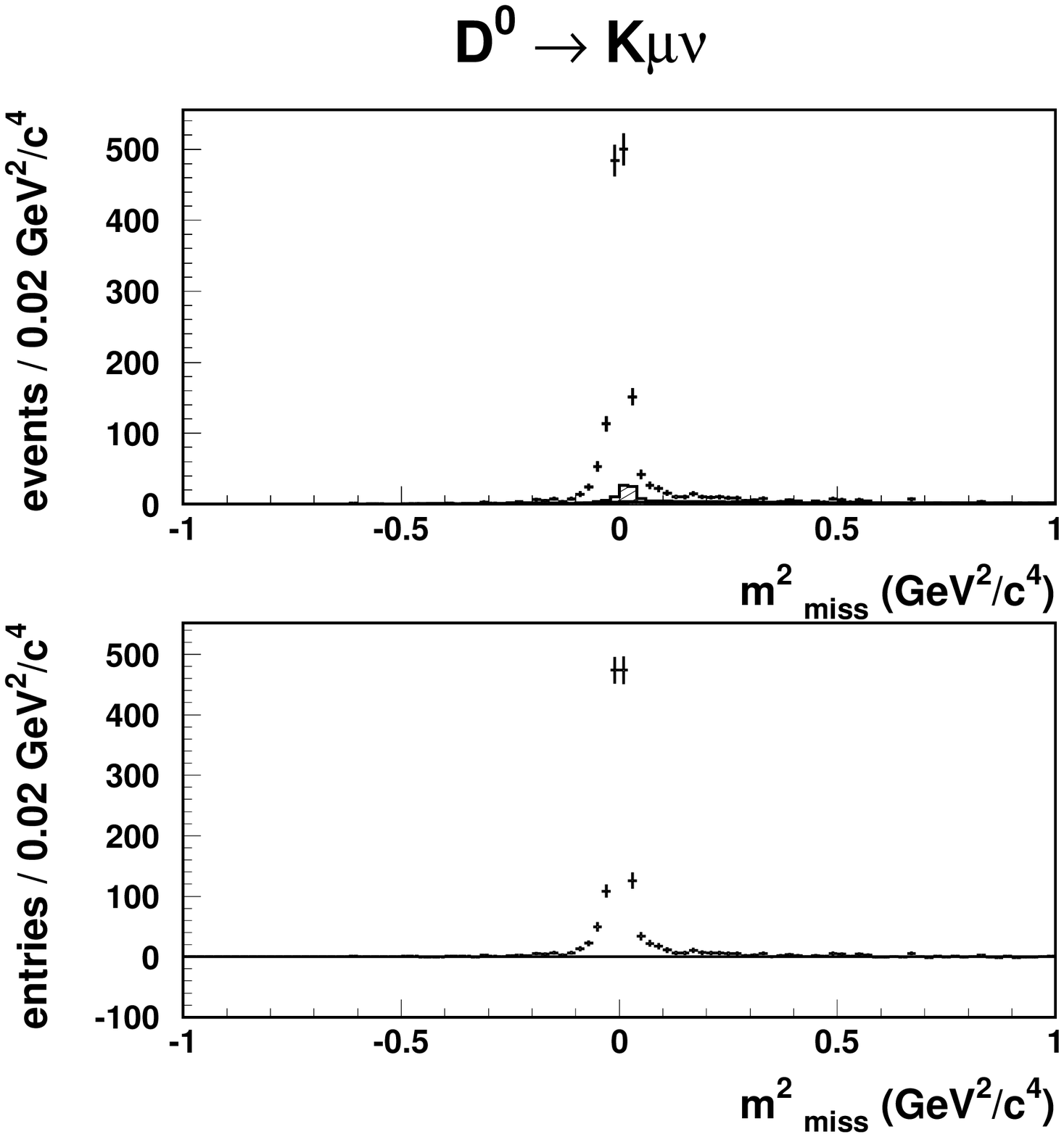}
    \includegraphics[width=8.0cm]{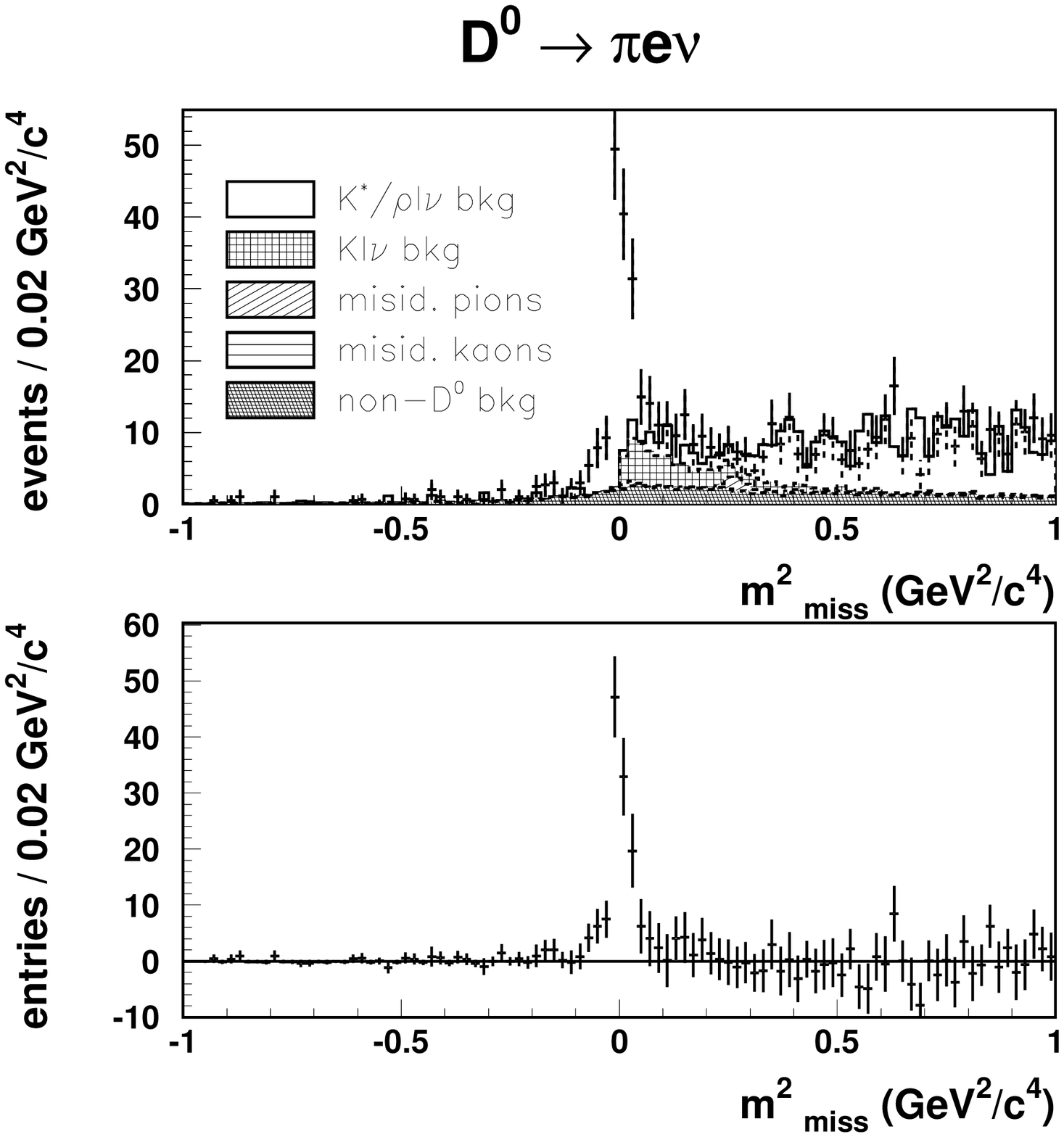}
    \includegraphics[width=8.0cm]{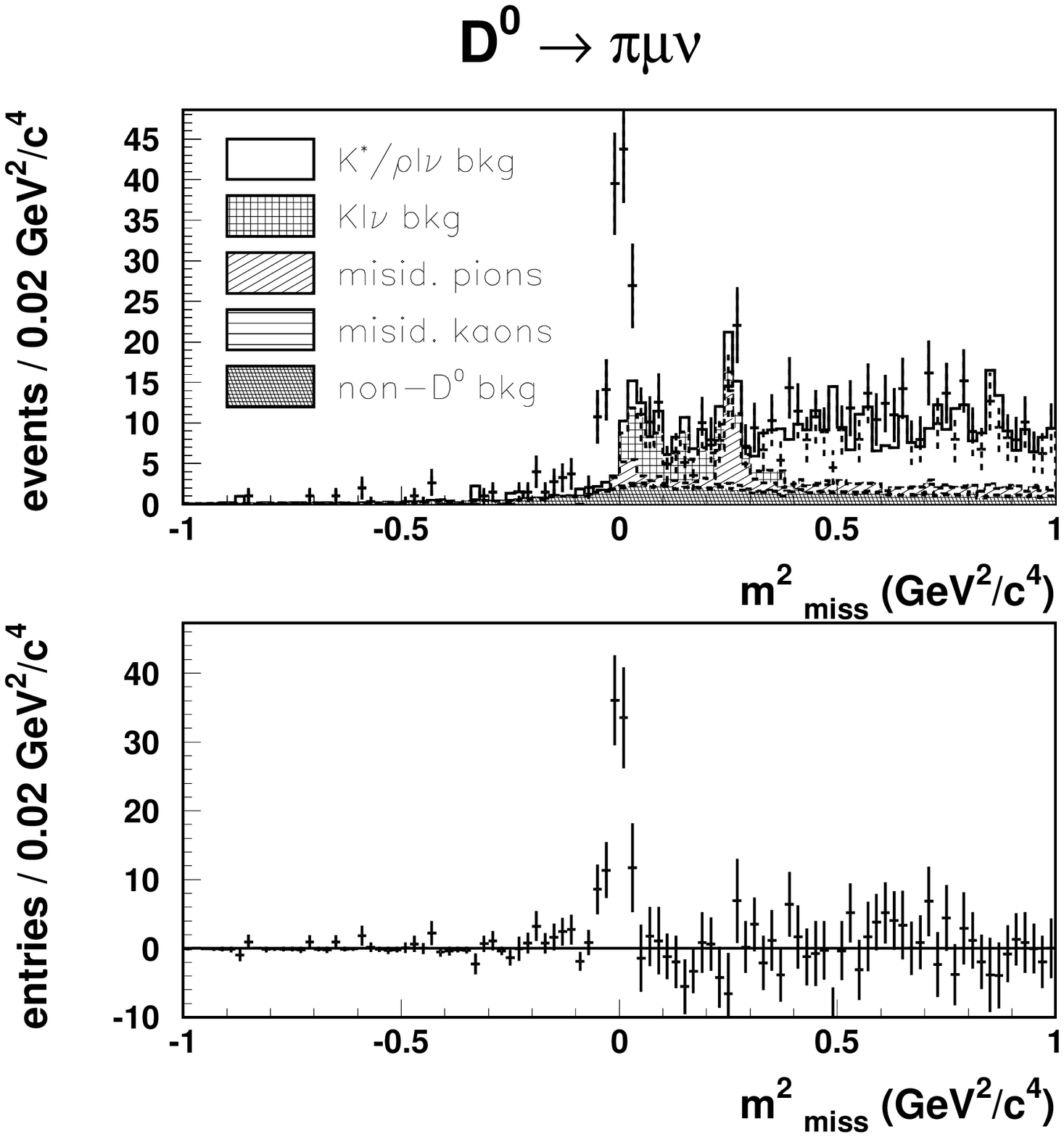}
  \end{center}
  \caption{Decomposition of background for the four semileptonic channels. In each case the lower plot shows the background-subtracted signal.}
\label{allbkg}
\end{figure}

Fig.~\ref{allbkg} shows the measured backgrounds and the signal 
for all four semileptonic decays of the $D^0$. Table  \ref{tab:events} gives 
the yields for the background sources.
The MC sample was used to check that all the significant background components are taken into account 
using the above described methods.

\section{Branching ratios of semileptonic decays of $D^0$}

After background subtraction, the yields for 
the four studied 
semileptonic $D^0$ decay modes are given in Table \ref{tab:events}.
The contributions to systematic uncertainties for each individual background 
source are listed in Table \ref{tab:syst}. 

\begin{table}[h]
\begin{center}
\begin{tabular}
{@{\hspace{0.5cm}}l@{\hspace{0.5cm}}||@{\hspace{0.5cm}}c@{\hspace{0.5cm}}|@{\hspace{0.5cm}}c@{\hspace{0.5cm}}|@{\hspace{0.5cm}}c@{\hspace{0.5cm}}|@{\hspace{0.5cm}}c@{\hspace{0.5cm}}}
  \hline
  channel & $K^-e^+\nu$ & $K^-\mu^+\nu$ & $\pi^- e^+\nu$ & $\pi^-\mu^+\nu$\\
  \hline
  \hline
  total yield & $1349$ & $1333$ & $152$ & $141$ \\
  \hline
  \hline
  signal yield & $1318$ & $1248$ & $126$ & $107$ \\
  \hline
  statistical error & 37 & 37 & 12 & 12 \\
  total systematic error & 7 & 24 & 2.6 & 6 \\
  \hline
  \hline
  non-$D^0$ background & $12.6\pm2.2$ & $12.4\pm4.8$ & $12.3\pm2.2$ & $11.5\pm4.5$ \\
  semileptonic background & $6.7\pm2.6$ & $10.0\pm2.5$ &  $11.7\pm1.2$ & $12.6\pm1.9$ \\ 
  hadronic background & $11.9\pm5.6$ & $62.1\pm23.9$ & $1.8\pm0.7$ & $9.7\pm3.7$ \\
  \hline
  \hline
\end{tabular}
\end{center}
 \caption{Final signal yields and backgrounds.}
 \label{tab:events}
\end{table} 

\begin{table}[h]
\begin{center}

\begin{tabular}
{@{\hspace{0.5cm}}l@{\hspace{0.5cm}}||@{\hspace{0.5cm}}c@{\hspace{0.5cm}}|@{\hspace{0.5cm}}c@{\hspace{0.5cm}}|@{\hspace{0.5cm}}c@{\hspace{0.5cm}}|@{\hspace{0.5cm}}c@{\hspace{0.5cm}}}
  \hline
  channel & $K^-e^+\nu$ & $K^-\mu^+\nu$ & $\pi^- e^+\nu$ & $\pi^-\mu^+\nu$\\
  \hline
  \hline
  {\bf non-$D^0$ background:} \\
  \hline
  statistics of background shape            & 4.2\%   & 8.1\%   & 6.6\%   & 13.4\% \\
  subtraction of wrong sign & 1.0\%   & 1.0\%   & 1.3\%   & 1.3\% \\
  charge correlated background            & 6.3\%   & 13.1\%   & 5.5\%   & 11.9\% \\
  shape minimum bias            & 15.7\%   & 35.2\%   & 15.3\%   & 34.1\% \\
  \hline
  sum in quadrature           & 17.5\%   & 38.4\%   & 17.6\%   & 38.5\% \\
  \hline
  {\bf semileptonic background:} \\
  \hline
  statistics MC samples              & 39\%   & 25\%   & 9.7\%   & 14.4\% \\
  fake rates ratio MC/data       & 4\%   & 3\%   & 2.9\%   & 3.2\% \\
  uncertainty of $K^{-*}\ell^+\nu/\rho^-\ell^+\nu$       & $< 1\%$   & $< 1\%$   & 1.4\%   & 1.7\% \\
  \hline
  sum in quadrature            & 39.2\%   & 25.2\%   & 10.2\%   & 14.8\% \\
  \hline
  {\bf hadronic background:} \\
  \hline
  fit of wrong sign shapes, mis-id kaons & 2\%   & 1\%   & 3\%   & 3\% \\
  fit of wrong sign shapes, mis-id pions & 33\%   & 16\%   & 35\%   & 23\% \\
  bias of background shapes & 33\%   & 35\%   & 11\%   & 30\% \\
  \hline
  sum in quadrature            & 46.7\%   & 38.5\%   & 36.8\%   & 37.9\% \\
  \hline
  \hline
\end{tabular}
\end{center}
 \caption{Relative systematic errors on the amount of respective backgrounds.}
 \label{tab:syst}
\end{table}

\begin{figure}
   \begin{center}
      \includegraphics[width=8.0cm]{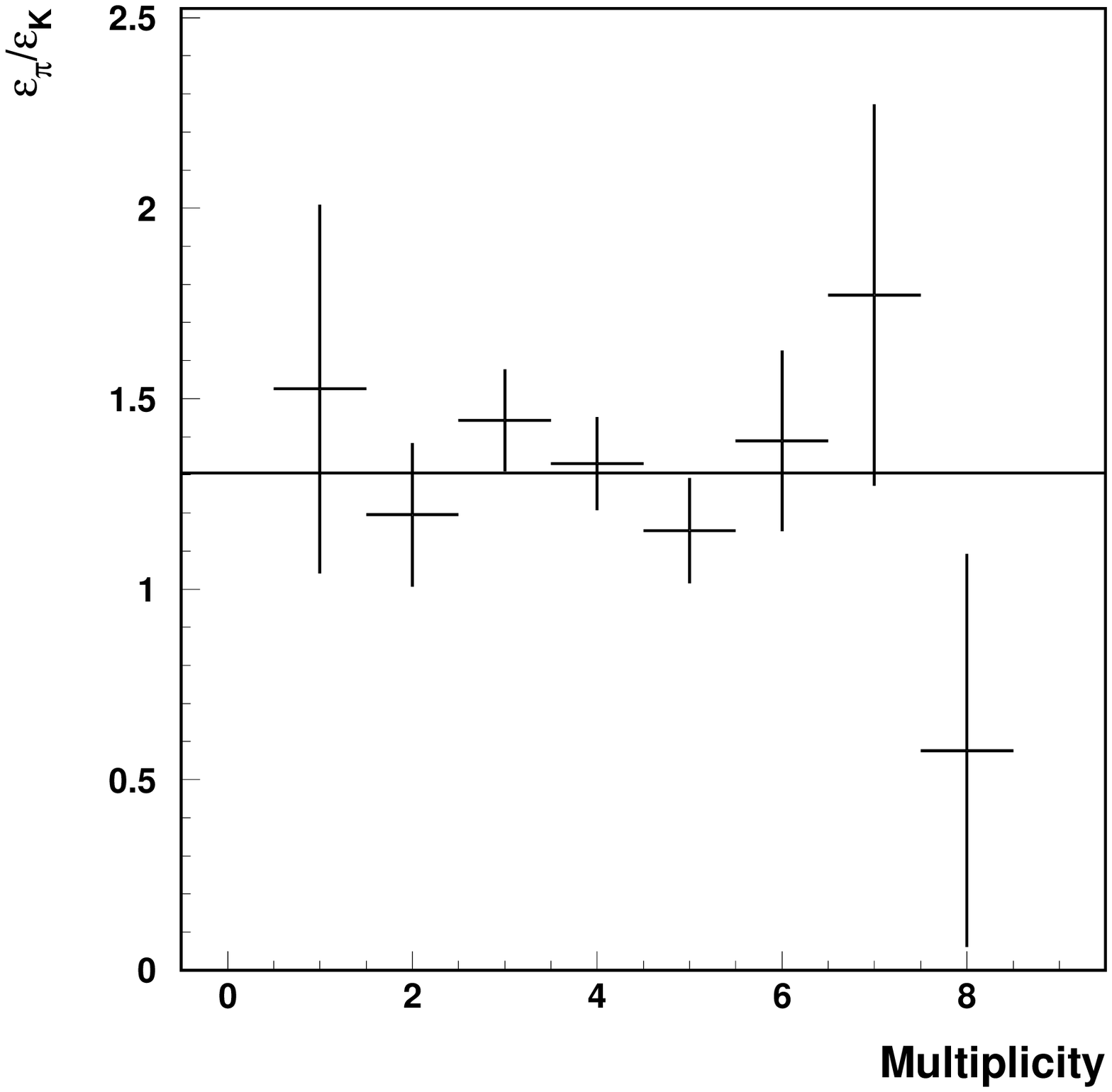}
   \end{center}
   \caption{Efficiency ratio for $D^0 \to \pi^-\ell^+\nu/D^0 \to K^-\ell^+\nu$ 
generic MC events as a function of multiplicity (number of additional mesons). The
solid line is the result of a fit to a constant.
   } \label{fig:5a}
 \end{figure}

The dependence of signal yields on the multiplicity has been studied
with generic MC.  
Although the efficiency decreases as the multiplicity of the event increases, the ratio 
$\epsilon_\pi/\epsilon_K$, where $\epsilon_{\pi}$ ($\epsilon_K$) is the efficiency of the $D^0 \to \pi^-\ell^+\nu$ ($K^-\ell^+\nu$) channel,
is flat (Fig. \ref{fig:5a}, $\chi^2/ndf = 0.84$).
The observed multiplicity
in the data is slightly different from the simulated one. In order to
estimate the systematic error due to this difference, we reweighted the
simulated efficiencies according to the data multiplicity distribution.
Since the $D^0 \to \pi^-\ell^+\nu$ and $D^0 \to K^-\ell^+\nu$ decay samples are topologically very
similar, we used a sample of events where only the selection on the
signal-side $D^0$ meson was applied. The efficiency ratio for the reweighted 
inclusive $D^0$ sample and the $D^0 \to K^-\ell^+\nu$ sample compared to the simulated
ratio yields a correction factor of $0.981 \pm 0.028$. Since this is compatible with unity, no correction is applied to the efficiency ratio; instead, the uncertainty of
this factor is included in the systematic error.
The values for the efficiencies,  corrected and averaged over multiplicity, are listed in Table \ref{tab:effs}.
\begin{table}
\begin{center}
\begin{tabular}
{@{\hspace{0.5cm}}l@{\hspace{0.5cm}}||@{\hspace{0.5cm}}c@{\hspace{0.5cm}}|@{\hspace{0.5cm}}c@{\hspace{0.5cm}}}
  \hline
  channel & $K^-e^+\nu$ & $K^-\mu^+\nu$ \\
  \hline
  \hline
  efficiency & $(0.369\pm0.005\pm0.010)\%$ & $(0.350\pm0.005\pm0.010)\%$ \\
  \hline
  \hline
  channel & $\pi^- e^+\nu$ & $\pi^-\mu^+\nu$\\
  \hline
  \hline
  efficiency & $(0.436\pm0.006\pm0.012)\%$ & $(0.443\pm0.006\pm0.012)\%$\\
  \hline
  \hline
\end{tabular}
\end{center}
 \caption{Efficiencies for 
semileptonic decay channels, averaged over multiplicity; errors shown are due to MC statistics and multiplicity dependence, as described in the text.}
 \label{tab:effs}
\end{table} 

Using these efficiencies, the relative 
branching ratios 
are obtained and are shown in Table \ref{tab:relbrs}. 
We have divided our data sample into low- and high-multiplicities sets of three or fewer and four or more pions (kaons), respectively, and found consistent results.
The results are also in agreement with recent measurements from CLEO \cite{ref:cleoff,ref:cleoc} 
and FOCUS \cite{ref:focus2} 

\begin{table}
\begin{center}
\begin{tabular}
{@{\hspace{0.1cm}}l@{\hspace{0.3cm}}||@{\hspace{0.3cm}}c@{\hspace{0.5cm}}|@{\hspace{0.5cm}}c@{\hspace{0.5cm}}|@{\hspace{0.5cm}}c@{\hspace{0.5cm}}}
  \hline
  relative BRs & this analysis & PDG \cite{ref:PDG} & theor. pred. \cite{ref:modpol}\\
  \hline
  \hline
   & & & \\
  $\frac{\mbox{BR}(D^0 \rightarrow K^- e^+ \nu)}{\mbox{BR}(D^0 \rightarrow K^- \mu^+ \nu)}$ & $1.002\pm0.041_{\mbox{stat}}\pm0.048_{\mbox{syst}}$ & $1.12\pm0.08$ & $1.03$ \\
   & &  &\\
  $\frac{\mbox{BR}(D^0 \rightarrow \pi^- e^+ \nu)}{\mbox{BR}(D^0 \rightarrow \pi^- \mu^+ \nu)}$ & $1.20\pm0.18_{\mbox{stat}}\pm0.09_{\mbox{syst}}$ &  & $1.02$ \\
   & &  &\\
  $\frac{\mbox{BR}(D^0 \rightarrow \pi^- e^+ \nu)}{\mbox{BR}(D^0 \rightarrow K^- e^+ \nu)}$ & $0.0809\pm0.0080_{\mbox{stat}}\pm0.0032_{\mbox{syst}}$ & $0.101\pm0.018$ & $0.086$ \\
   & &  &\\
  $\frac{\mbox{BR}(D^0 \rightarrow \pi^- \mu^+ \nu)}{\mbox{BR}(D^0 \rightarrow K^- \mu^+ \nu)}$ & $0.0677\pm0.0078_{\mbox{stat}}\pm0.0047_{\mbox{syst}}$ &  & $0.087$ \\
   & &  &\\
  \hline
  \hline
\end{tabular}
\end{center}
 \caption{Relative branching ratios}
 \label{tab:relbrs}
\end{table} 

\section{$q^2$-distribution} \label{sec:q2dist}

\begin{figure}
  \begin{center}
    \includegraphics[width=7.cm]{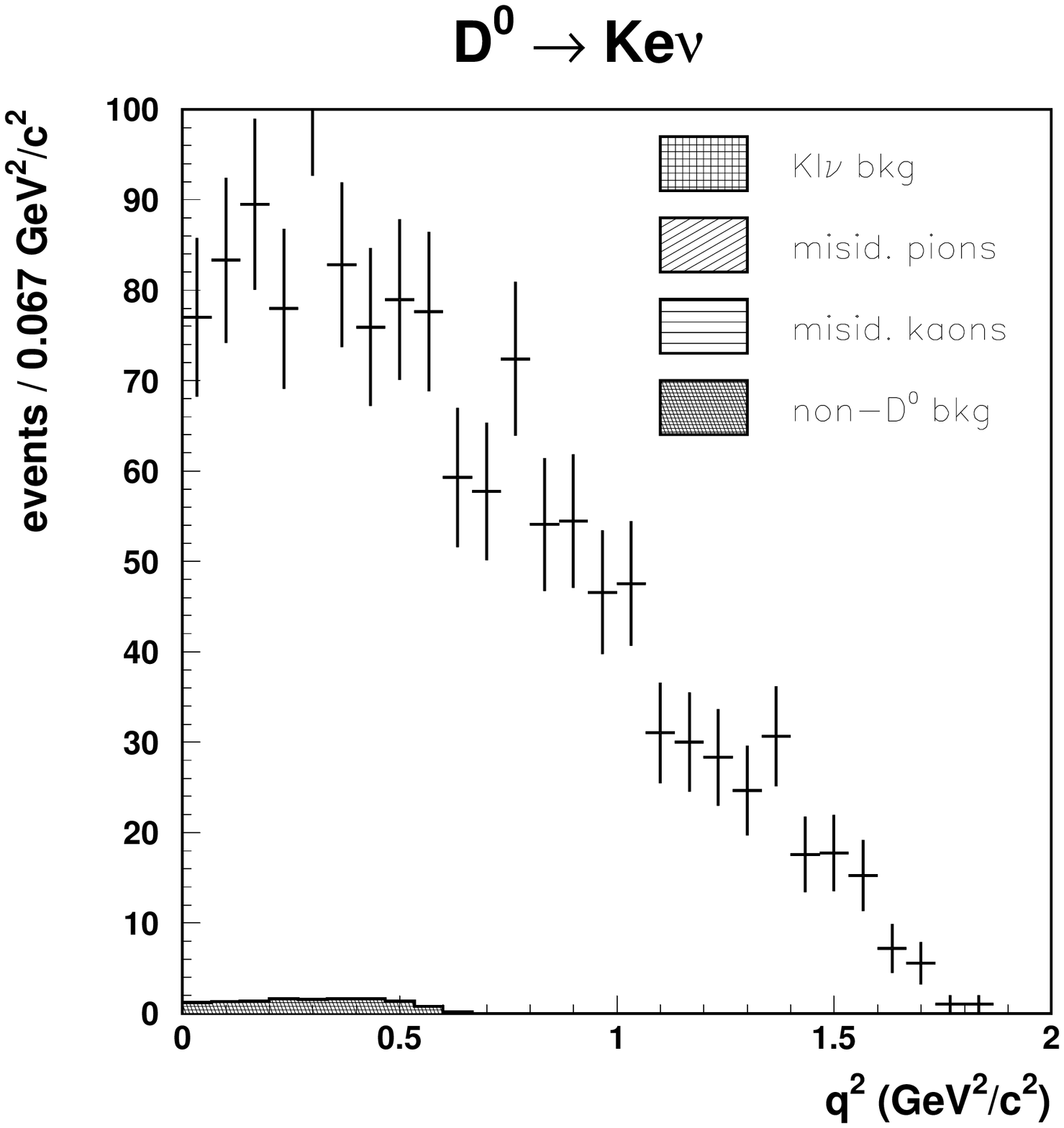}
    \includegraphics[width=7.cm]{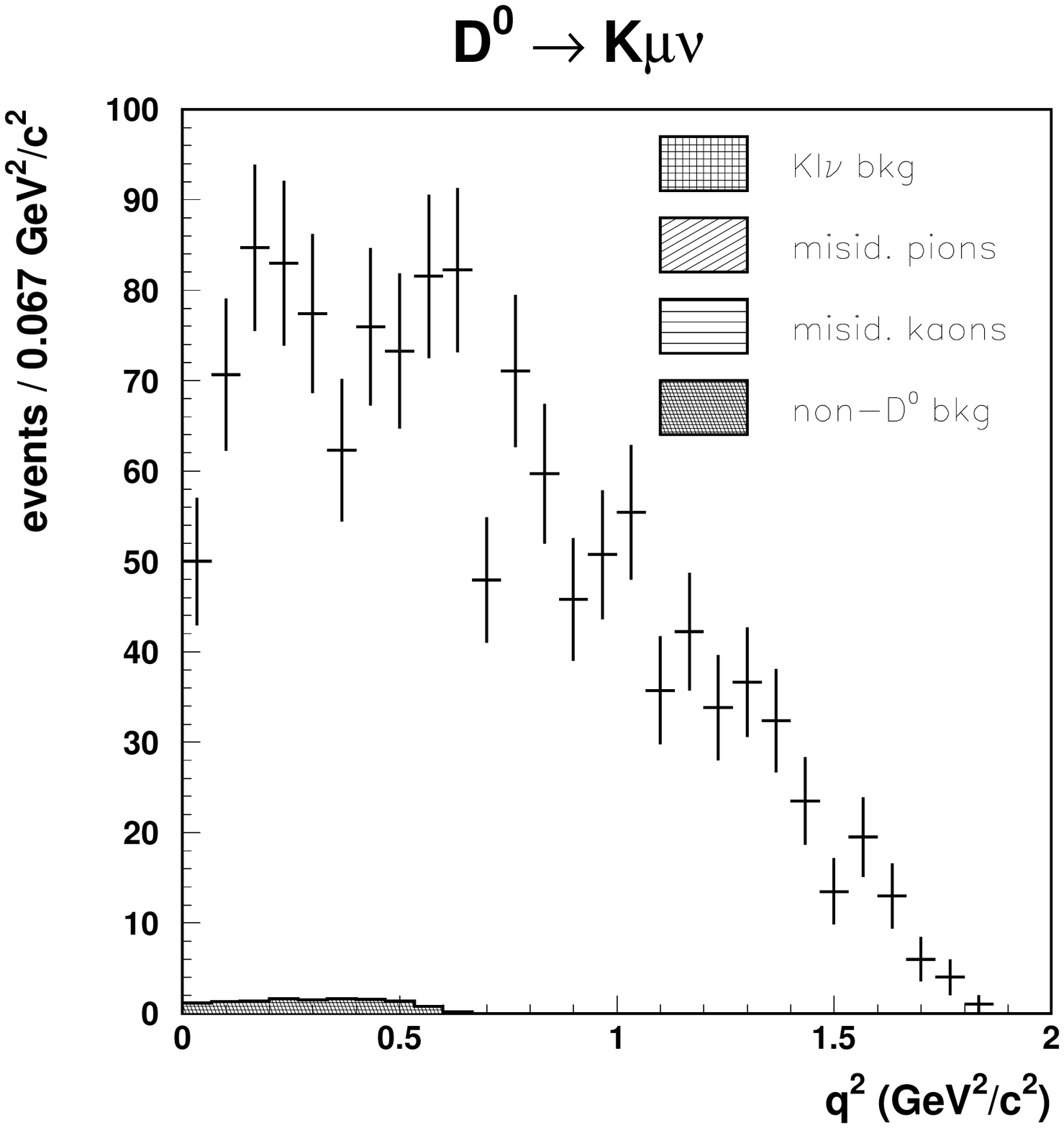} \\
    \includegraphics[width=7.cm]{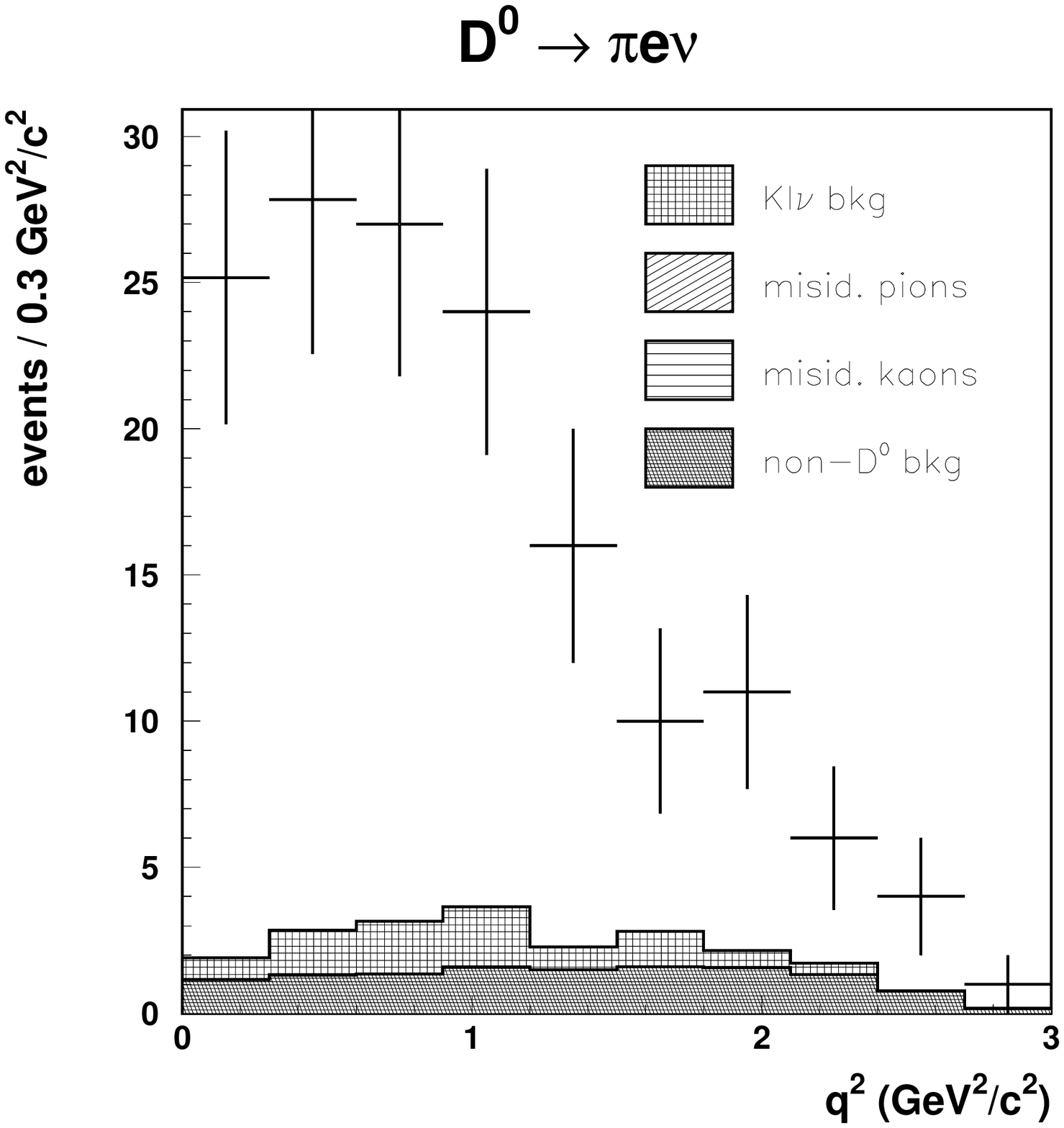} 
    \includegraphics[width=7.cm]{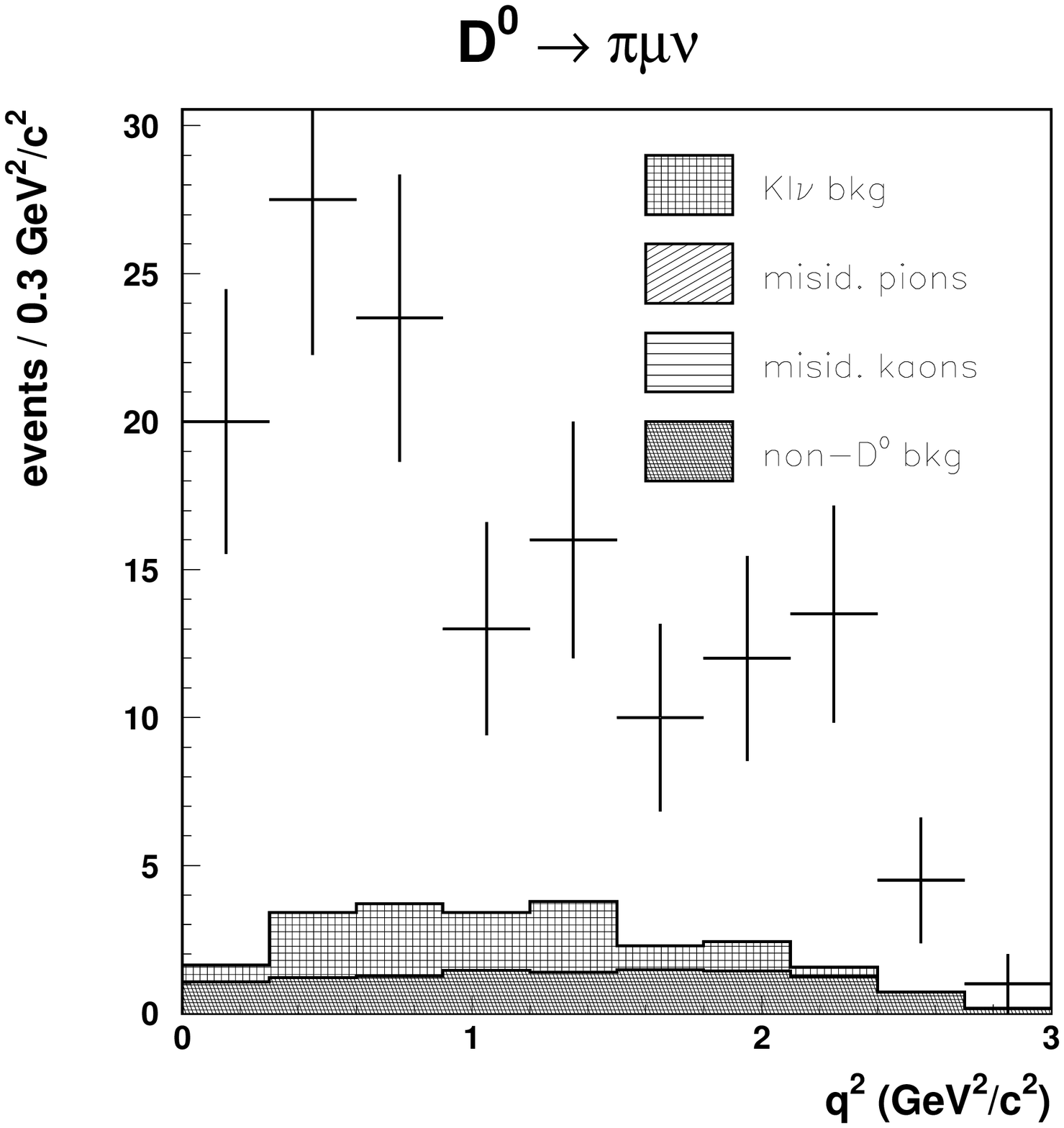}
  \end{center}
  \caption{$q^2$ distributions for $D^0 \to K^- e^+\nu$ (upper left), $D^0 \to K^-\mu^+\nu$ (upper right), $D^0 \to \pi^- e^+\nu$ (lower left) and $D^0 \to \pi^-\mu^+\nu$ (lower right), with contributions of individual background sources shown as histograms.}
\label{fig:q2dist}
\end{figure}

The square of the four-momentum transfer $q$ in the semileptonic channel is given by 
$q^2= (p_{\mathrm \nu}+p_{\mathrm \ell})^2$, 
where  $p$ is the four momentum of the specified particle. The good neutrino resolution leads also to a very 
good resolution for  $q^2$, described by a double Gaussian with a common mean of $0.00061 \pm 0.00031$ GeV$^2/c^2$ and widths of 
$\sigma_1 = 0.0109 \pm 0.0005 $ GeV$^2/c^2$, $\sigma_2 = 0.0401 \pm 0.0026 $ GeV$^2/c^2$, 
the fraction of the wider Gaussian being $0.14\pm 0.02$.
No unfolding of  the $q^2$ distributions is needed since these
resolutions are much less than the bin widths used. 
The dependence of $q^2$ resolution on different multiplicities 
is small; $\sigma(q^2)$ lies in the range $0.0095$ to
$0.0121$ GeV$^2/c^2$ (with a $\sim$ 10\% uncertainty) for pion/kaon multiplicities between $0$ and $8$.
The background subtraction in the determination of the $q^2$ distribution 
is performed using the method described earlier.
The results are shown in Fig. \ref{fig:q2dist} and are consistent with MC.

\section{\boldmath Semileptonic Form Factor for $D^0 \to \pi^-\ell^+\nu$ and $D^0 \to K^-\ell^+\nu$}

In general, for P $\rightarrow$ P transitions (P denoting a pseudo-scalar meson), two form factors 
$f_+(q^2)$ and $f_-(q^2)$ are needed
\cite{ref:2}.
In the case of semileptonic decays, the matrix element is dominated by 
the form factor $f_+$; the influence of $f_-$ is suppressed kinematically 
to a negligible (small) level for electrons (muons)
by the smallness of $m^2_l$ compared to the scales of $m^2_D$, $p^2_\pi$ and $q^2$.
Up to order $m_{\mathrm \ell}^2$ one has
\begin{eqnarray}
\frac{d\Gamma(D^0\rightarrow \pi^-\ell^+\nu)}{dq^2} & = & \frac{G_F^2|V_{cd}|^2}{24\pi^3}{|f_+(q^2)|^2}p_{\mathrm \pi}^3
\label{equ:ff}
\end{eqnarray}
and an analogous expression for $D^0 \to K^-\ell^+\nu$, where $p_{\mathrm \pi}$ is the magnitude  of the pion 3-momentum in the $D^0$ rest frame \footnote{For our analysis, 
we use the full formula of Ref. \cite{ref:2}, including $m^2_l$ terms, which gives a small correction 
for low $q^2$ values.}, determining the {\it kinematical weight} of the form factor.
These form factors have been calculated recently in unquenched lattice 
QCD \cite{ref:unquenched}, \cite{ref:unquenched2}. Earlier calculations were done in 
quenched lattice QCD~\cite{ref:1}, which works best in the region of high $q^2$ 
where measurements unfortunately suffer from poor statistics. 
In the {\it pole model} \cite{ref:2}, the form factor $f_+$ is described as
\begin{eqnarray}
  f_+(q^2) &= & \frac{f_+(0)}{(1-q^2/m_{\mbox{pole}}^2)}
\end{eqnarray}
with the pole masses 
$m_{\mathrm D_s^*} = 2.11$ GeV/$c^2$ (for $D^0 \rightarrow K^- \ell^+ \nu$) and
$m_{\mathrm D^*} = 2.01$ GeV/$c^2$ (for $D^0 \rightarrow \pi^- \ell^+ \nu$). 

Within the {\it modified pole model}~\cite{ref:modpol}, the form factor is given by
\begin{eqnarray}
  f_+(q^2) &= & \frac{f_+(0)}{(1-q^2/m_{\mbox{pole}}^2)(1-\alpha_p q^2/m_{\mbox{pole}}^2)}.
\end{eqnarray}

The ISGW2-model \cite{ref:3} predicts the following expression for the form factor:
\begin{eqnarray}
f_+(q^2) &= & \frac{f_+(q^2_{max})}{(1-\alpha_I (q^2-q^2_{max}))^2} = \frac{f_+(0)(1+\alpha_I q^2_{max})^2}{(1-\alpha_I (q^2-q^2_{max}))^2}
 \label{equ:isgw2}
\end{eqnarray}
 where $q^2_{max}$ is the kinematical limit of $q^2$ and $\alpha_I$ is a 
parameter of the model.

\begin{figure}
  \begin{center}
    \includegraphics[width=7.0cm]{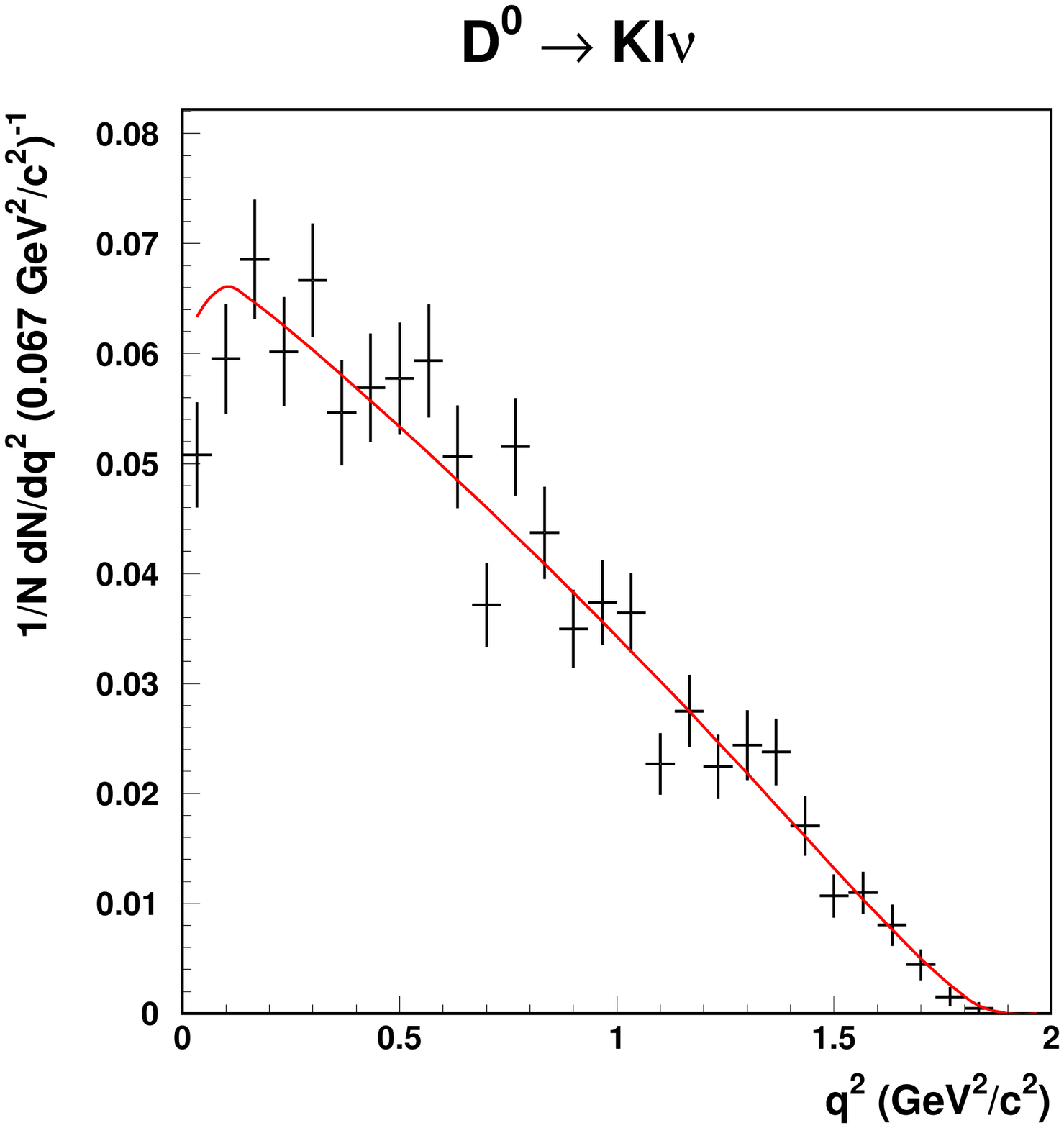}
    \includegraphics[width=7.0cm]{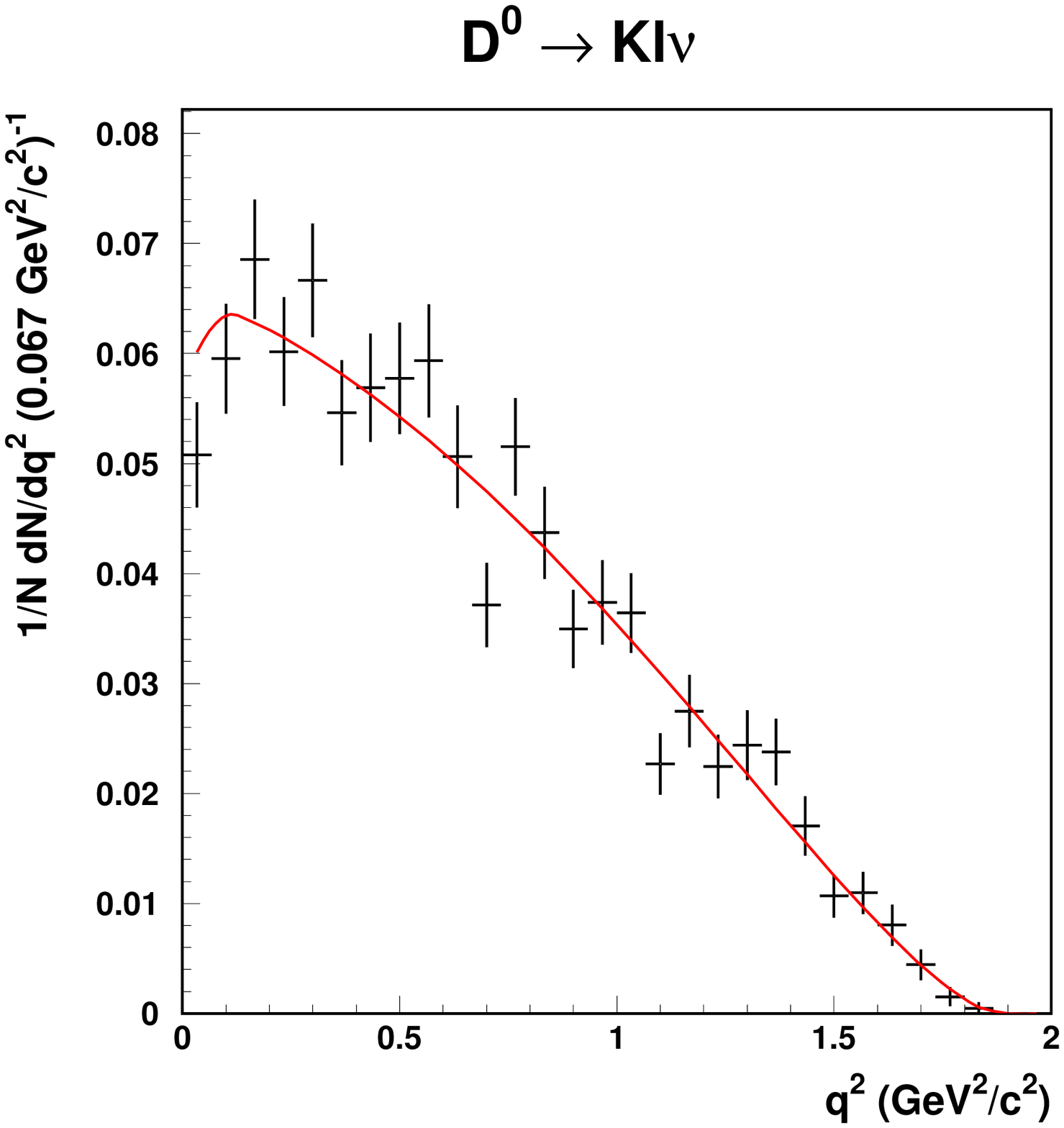}
    \includegraphics[width=7.0cm]{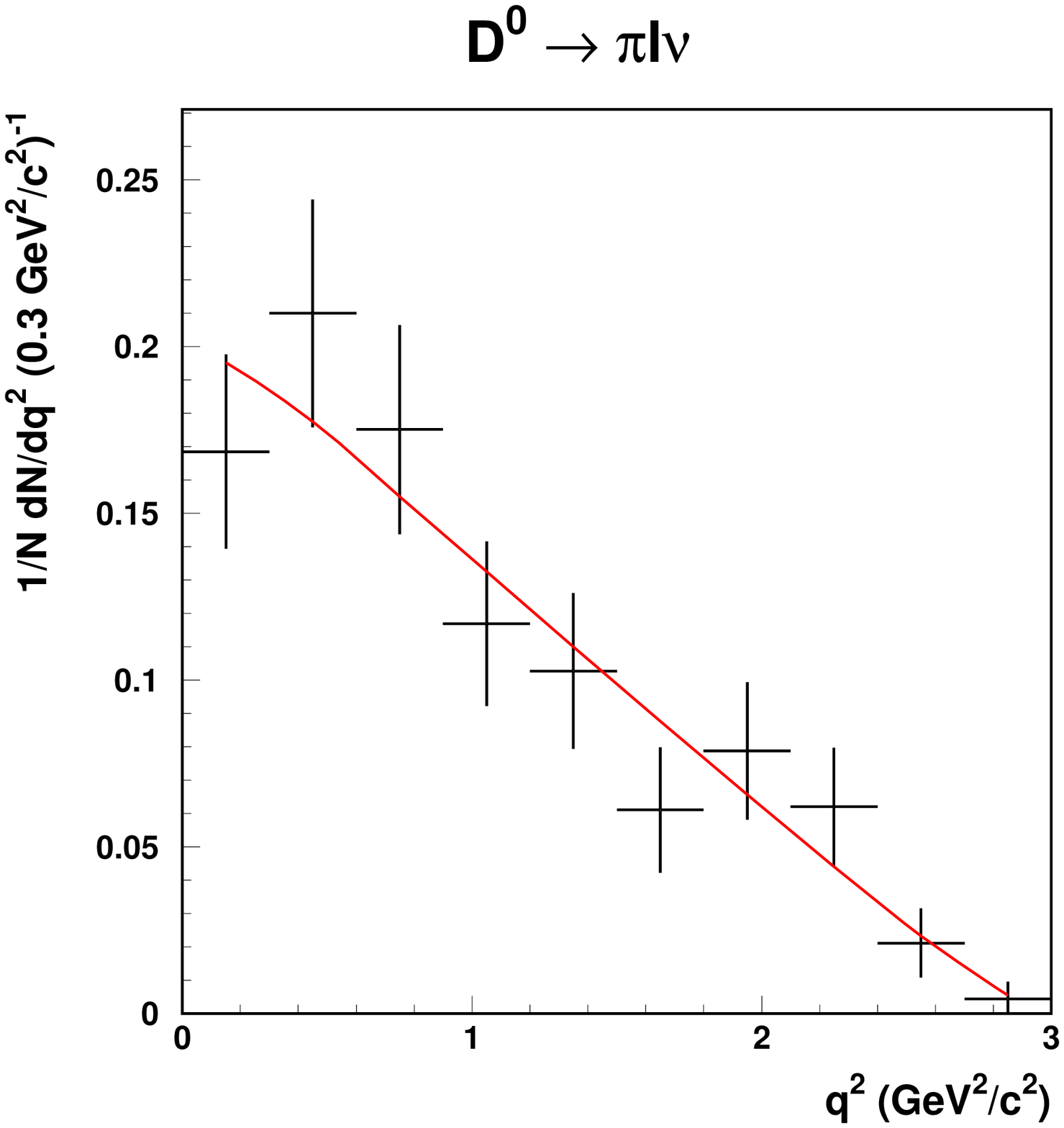}
    \includegraphics[width=7.0cm]{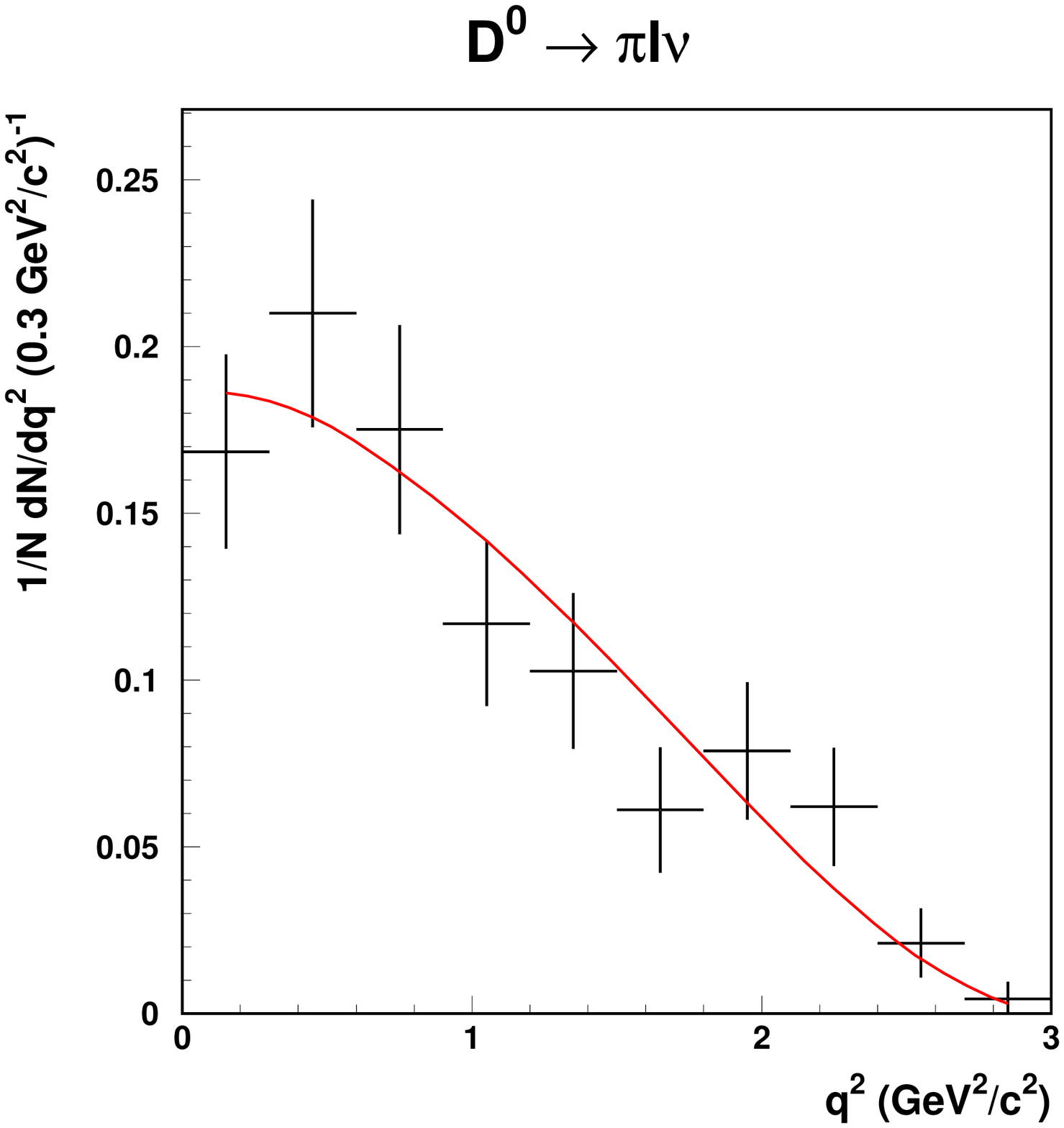}
  \end{center}
  \caption{Fraction of events in different $q^2$-bins for $D^0 \to  K^- \ell^+ \nu$ and $D^0 \to \pi^- \ell^+ \nu$ 
  compared with the fits of the simple pole model (left) and ISGW2 (right)}
\label{fig:11}
\end{figure}

The normalized, bin-by-bin background subtracted and efficiency
corrected $q^2$ distributions are shown in Fig. \ref{fig:11} for different decay modes; error bars indicate the sum in quadrature of statistical and systematic uncertainties.
We fit the measured distributions to the predicted normalized
differential decay width $\displaystyle\frac{1}{\Gamma}\frac{d\Gamma}{dq^2}$ of different models
described above. Binning effects are accounted for by averaging the
model functions over individual $q^2$ bins. Parameters $f(0)$ and $f(q_{max}^2)$
cancel out in the normalized decay width.

The fit results for the simple pole model with the pole mass as free parameter are shown in Fig. \ref{fig:11}, left. The fitted pole masses are
\begin{eqnarray}
m_{pole}(D^0 \to K^- e^+ \nu) & = & 1.88\pm0.06\pm0.03 \mbox{ GeV}/c^2 \\
m_{pole}(D^0 \to K^- \mu^+ \nu) & = & 1.77\pm0.04\pm0.03 \mbox{ GeV}/c^2 \\
m_{pole}(D^0 \to \pi^- e^+ \nu) & = & 2.01\pm0.13\pm0.04 \mbox{ GeV}/c^2 \\
m_{pole}(D^0 \to \pi^- \mu^+ \nu) & = & 1.92\pm0.09\pm0.04 \mbox{ GeV}/c^2 
\end{eqnarray}
where the first error is statistical and second systematic (discussed
below). The $\chi^2$/ndf values for fits using different models are shown in Table \ref{tab:chifit}.
Values of $m_{pole}(K \ell \nu)$ are several standard deviations below the $m_{D_s^*}$. The pole mass for the $\pi \ell \nu$ decay agrees within errors with the predicted value, $m_{D^*}$.
The fitted pole masses are also in agreement with results from 
CLEO \cite{ref:cleoff} and FOCUS \cite{ref:focus}.

A fit to the modified pole model is also performed,  
where $m_p$ is fixed to the theoretical pole $m_{D^*_{(s)}}$. The results for the fit parameter $\alpha_p$ are:
\begin{eqnarray}
\alpha_p (D^0 \to K^- e^+ \nu) & = & 0.40 \pm 0.12 \pm 0.09 \label{equ:alphap1}  \\
\alpha_p (D^0 \to K^- \mu^+ \nu) & = & 0.66 \pm 0.11  \pm 0.09  \\
\alpha_p (D^0 \to \pi^- e^+ \nu) & = & 0.03 \pm 0.27  \pm 0.13  \\
\alpha_p (D^0 \to \pi^- \mu^+ \nu) & = & 0.19 \pm 0.32  \pm 0.16  \label{equ:alphap4} 
\end{eqnarray}
Significantly non-zero values of $\alpha_p(K \ell \nu)$ suggest further
contributions to the form factor, apart from the single pole at $m_{D_s^*}$.

Finally, a fit of the parameter $\alpha_I$ in the ISGW2 model yields (Fig. \ref{fig:11}, right)
\begin{eqnarray}
\alpha_I (D^0 \to K^- e^+ \nu) & = & 0.37 \pm 0.03  \pm 0.02  \mbox{ GeV}^{-2}c^2  \\
\alpha_I (D^0 \to K^- \mu^+ \nu) & = & 0.44 \pm 0.03  \pm 0.02  \mbox{ GeV}^{-2}c^2  \\
\alpha_I (D^0 \to \pi^- e^+ \nu) & = & 0.35 \pm 0.08  \pm 0.04   \mbox{ GeV}^{-2}c^2 \\
\alpha_I (D^0 \to \pi^- \mu^+ \nu) & = & 0.40 \pm 0.10  \pm 0.03   \mbox{ GeV}^{-2}c^2 . 
\end{eqnarray}
The $\chi^2$ of the fit to the ISGW2 model is comparable to that for the
two other models (c.f. Table \ref{tab:chifit}). The prediction of \cite{ref:3} for the $K\ell\nu$ mode is $\alpha_I = 0.47$, which is within one standard deviation of the $K \mu \nu$ result, and almost three standard deviations from the $K e \nu$ result.

\begin{table}
\begin{center}
\begin{tabular}
{@{\hspace{0.5cm}}l@{\hspace{0.5cm}}||@{\hspace{0.5cm}}c@{\hspace{0.5cm}}|@{\hspace{0.5cm}}c@{\hspace{0.5cm}}|@{\hspace{0.5cm}}c@{\hspace{0.5cm}}|@{\hspace{0.5cm}}c@{\hspace{0.5cm}}}
  \hline
  $\chi^2/ndf$ & $K^-e^+\nu$ & $K^-\mu^+\nu$ & $\pi^- e^+\nu$ & $\pi^-\mu^+\nu$\\
  \hline
  \hline
  simple pole model & $1.08$ & $1.40$ & $0.37$ & $0.85$ \\
  modified pole model & $1.05$ & $1.35$ & $0.37$ & $0.89$ \\
  ISGW2 & $1.02$ & $1.33$ & $0.35$ & $1.09$ \\
  \hline
  \hline
\end{tabular}
\end{center}
 \caption{$\chi^2$ per degree of freedom for fits to various models}
 \label{tab:chifit}
\end{table} 
To study systematic uncertainties in different fits, each fit was
 repeated 50000 times, with the individual background normalizations    
 varied within their Gaussian errors in a correlated manner for all $q^2$  
 bins. The mean differences from the results of default fits were taken 
 as systematic errors. A similar procedure was adopted to estimate the
 error due to the uncertainty on the $q^2$ dependent efficiency. 
The systematic error is included in the fit results above.

\begin{figure}
  \begin{center}
    \includegraphics[width=7.0cm]{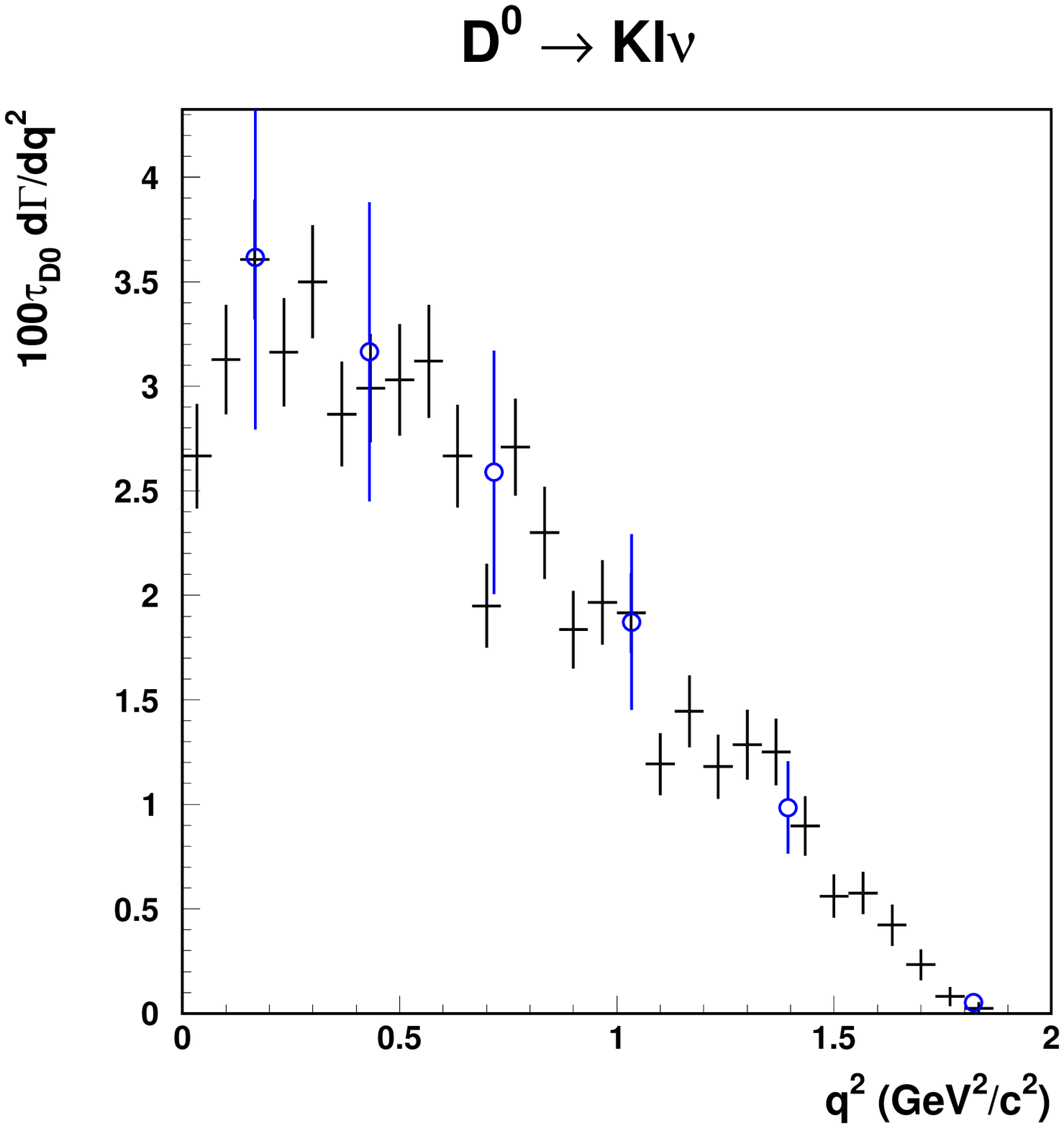}
    \includegraphics[width=7.0cm]{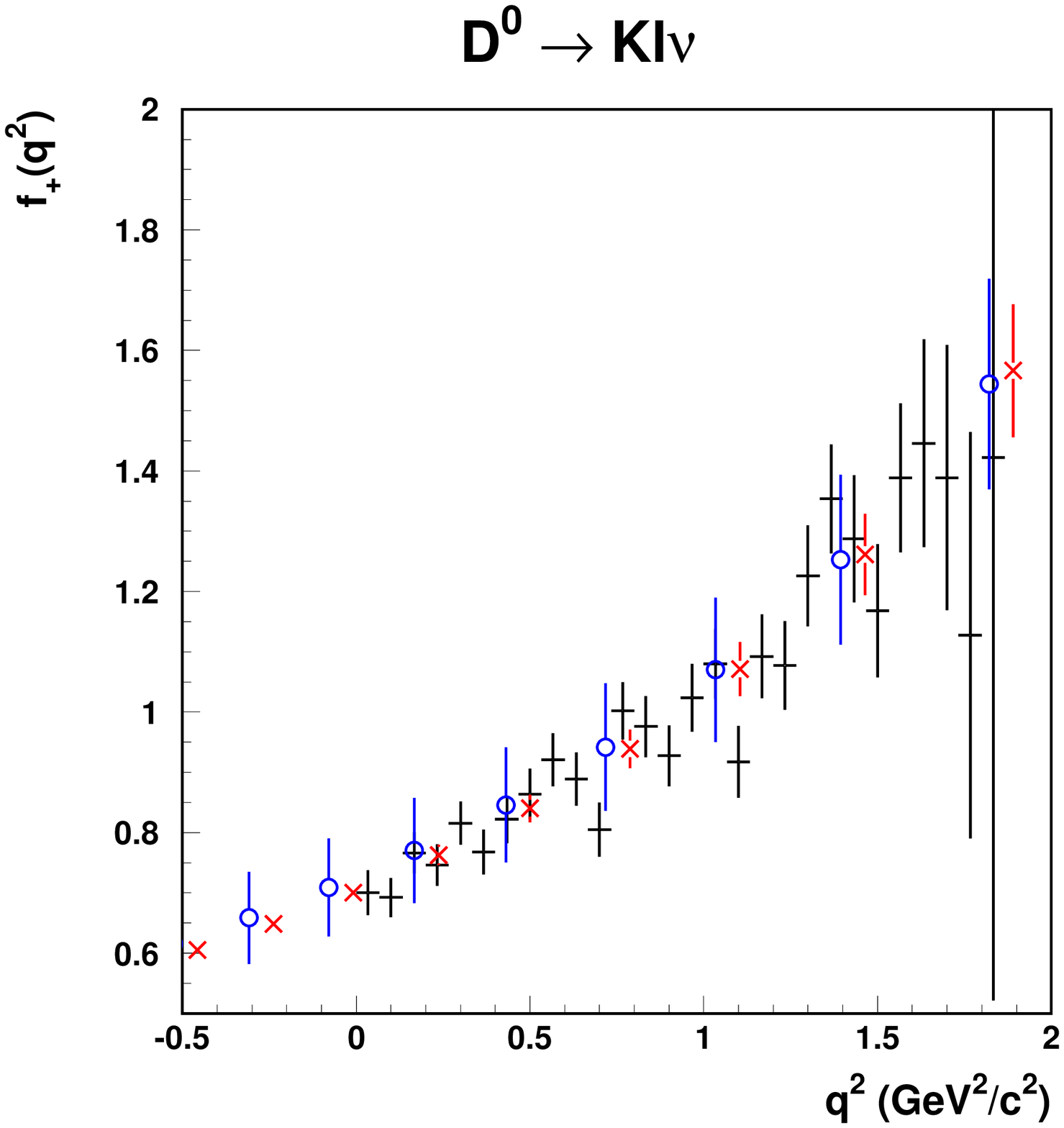}
    \includegraphics[width=7.0cm]{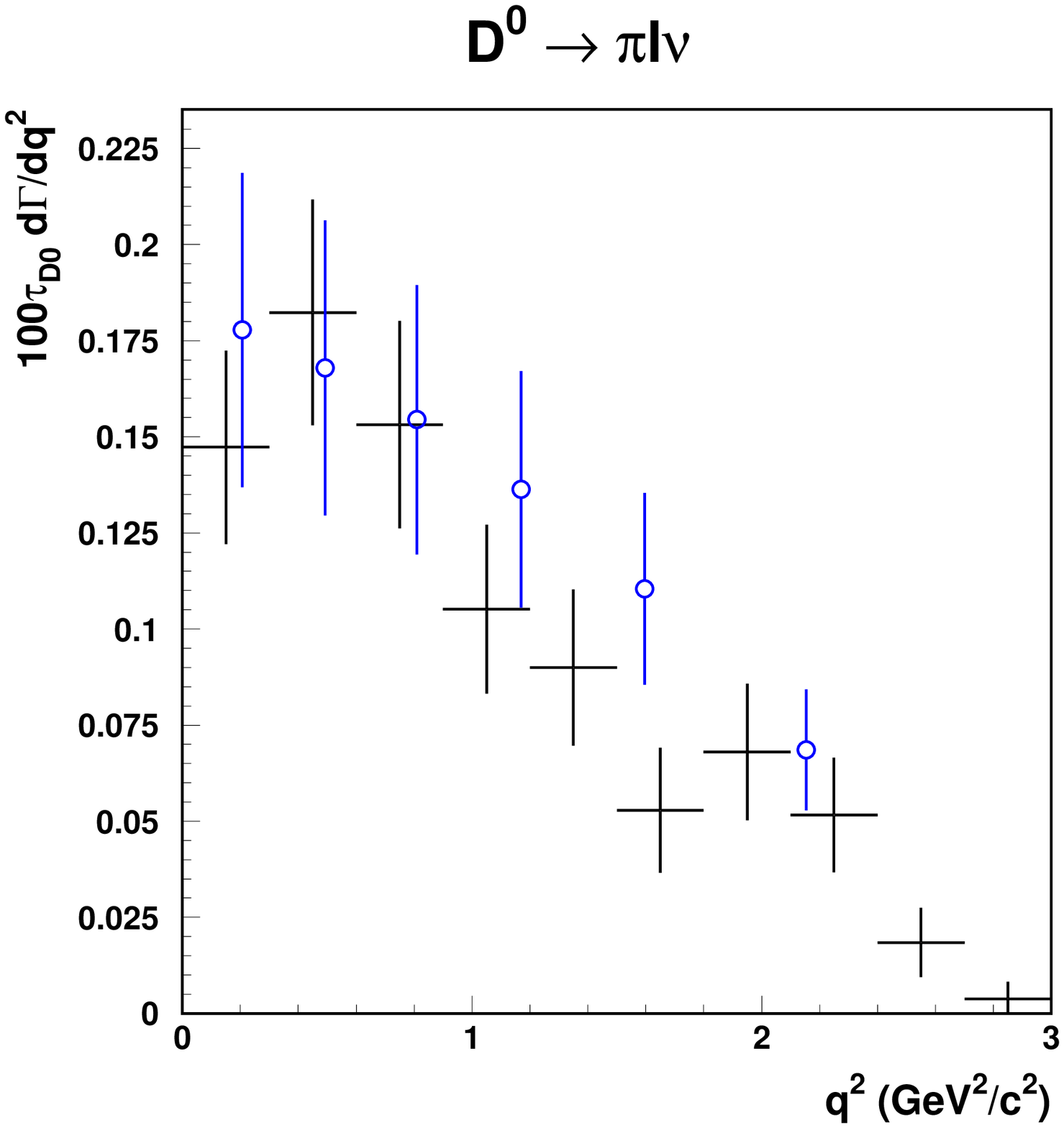}
    \includegraphics[width=7.0cm]{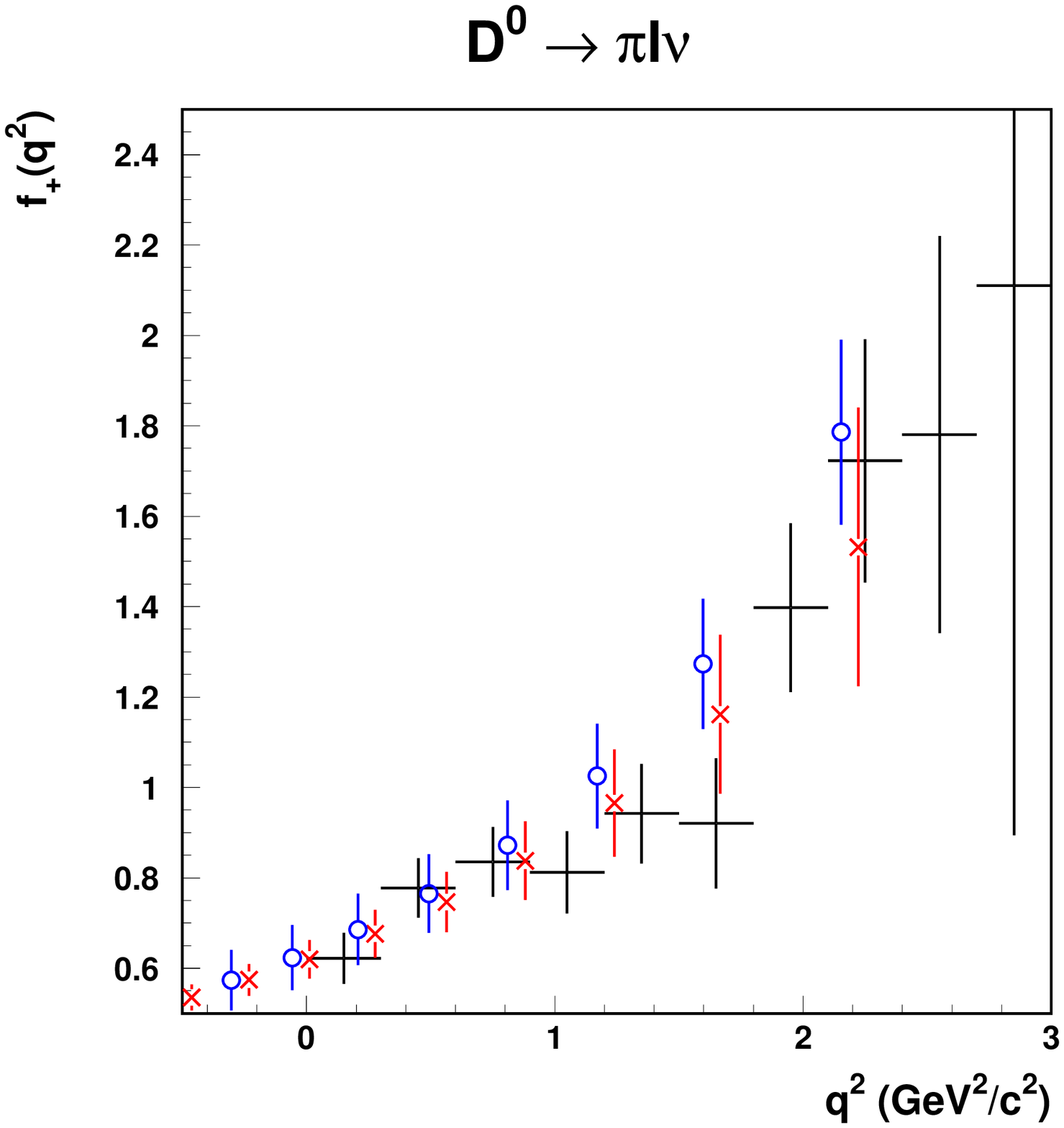}
  \end{center}
  \caption{
Left plot: comparison of the absolute partial decay width $d\Gamma/dq^2$ with a lattice calculation (circles);
right plot: 
comparison of the measured form factor (error bars) and
unquenched lattice results (circles). To make a more direct comparison,
$f_+(q^2)$ is also calculated at the same $q^2$ values where the lattice
prediction is available (shifted
slightly higher to avoid overlap), using the modified pole model  with $\alpha_p$
parameter from the fit to the data (crosses)}
 \label{fig:X}
\end{figure}

The absolute (i.e. not normalized) partial decay width is related to observable quantities by
\begin{equation}
\frac{d\Gamma(D^0 \rightarrow (\pi/K)^-\ell^+\nu)}{dq^2}=\frac{\epsilon_K BR_K N(q^2)}{\tau_D \epsilon_{q^2} N_K} \label{equ:pdw}
\end{equation}
where $\epsilon_K$ , $BR_K\,$ and $N_K$ are the average efficiency, branching fraction, and number of reconstructed signal events in the $D^0 \to K^-\ell^+\nu$ mode, respectively. The world average value \cite{ref:PDG} is used for the branching fraction, as well as for the $D^0$ lifetime $\tau_D$. $\epsilon_{q^2}$ and $N(q^2)$ are the efficiencies and signal yields in individual $q^2$ bins.
A comparison of the measured values
with lattice calculations \cite{ref:unquenched} is shown in Fig. \ref{fig:X} (left).
The uncertainty of the lattice results is mainly
systematic (errors given in \cite{ref:unquenched} are statistical only; relative
systematic errors of about $10\%$ were added in quadrature according to \cite{ref:okamotopriv}).
The lattice results in Ref.~\cite{ref:unquenched2} are very similar to those in Ref.~\cite{ref:unquenched}, and are therefore not shown here. 

From the absolute partial decay width, also the absolute form factor can be extracted using Eqn. \ref{equ:ff}. 
Comparison of the measured values and lattice calculations \cite{ref:unquenched} is shown
in Fig. \ref{fig:X} (right). In order to make a more direct comparison, $f_+(q^2)$ is
also calculated at the same $q^2$ values where the prediction is
available. For interpolation we use the modified pole model (also used
in \cite{ref:unquenched}) with $\alpha_p$ obtained from the fit to data described above.

Using the integrated form of Eqn. \ref{equ:ff} we determine the ratio of the form
factors for $K \ell \nu$ and $\pi \ell \nu$ decay modes at $q^2=0$. For the $q^2$
dependence of the form factor we use the modified pole model with the
values of parameter $\alpha_p$ in Eqn. \ref{equ:alphap1}-\ref{equ:alphap4}, with the weighted average of $K$
and $\pi$ modes calculated taking into account correlated systematic
errors. A similar averaging procedure is done for the measured ratios of
branching fractions (Table \ref{tab:relbrs}). We measure the ratio
\begin{eqnarray}
\left. \frac {f_+({D^0 \to \pi^- \ell^+ \nu})^2|V_{cd}|^2} {f_+({D^0 \to K^- \ell^+ \nu})^2|V_{cs}|^2}\right\arrowvert_{q^2=0} & = & 0.041\pm 0.003_{\mbox{stat}}\pm 0.004_{\mbox{syst}} 
\end{eqnarray}
which is consistent within errors with the model-independent result using only the data in the first $\pi\ell\nu$ $q^2$ bin ($q^2<0.3\mathrm{\ GeV}^2/c^2$). A recent theoretical prediction for the ratio~\cite{ref:unquenched} is $0.040\pm0.004$.

This result is in fair agreement with those from CLEO \cite{ref:cleoff} and FOCUS \cite{ref:focus2}, which measure slightly lower values.

\section{Summary and conclusions}

In the present measurement we use fully reconstructed $e^+e^- \rightarrow D^{(*)}_{tag}D^{*\pm}_{signal}nh$ events ($h=\pi^{\pm 0}, K^{\pm}$), 
where the tag-side decay channels were $D^* \rightarrow D\pi, D\gamma$ and $D \rightarrow K^{\pm}n\pi$, $n=1-3$ and the signal side was reconstructed from the recoil as $D^{*\pm} \rightarrow D^0\pi^\pm$ and $D^0 \rightarrow K/\pi^-\ell^+\nu$, $l=e,\mu$. 
Mass constraints at each reconstruction step enabled a very good resolution for the momentum transfer $q^2$ of $\sigma_{q^2} \sim 0.01~GeV^2/c^2$.

We determined relative branching fractions 
\begin{eqnarray}
  \frac{\mbox{BR}(D^0 \rightarrow \pi^- e^+ \nu)}{\mbox{BR}(D^0 \rightarrow K^- e^+ \nu)} & = & 0.0809\pm0.0080_{\mbox{stat}}\pm0.0032_{\mbox{syst}} \\
  \frac{\mbox{BR}(D^0 \rightarrow \pi^- \mu^+ \nu)}{\mbox{BR}(D^0 \rightarrow K^- \mu^+ \nu)} & = & 0.0677\pm0.0078_{\mbox{stat}}\pm0.0047_{\mbox{syst}} 
\end{eqnarray}
in good agreement with other measurements.
The normalized measured $q^2$ distribution was fitted to different models
of form factors. In the $K \ell \nu$ decay mode, where the sample of events
with higher statistics is selected, some significant deviations from the
predictions of the simple pole and ISGW2 model are observed. The results
of the fit to the pole masses are
$m_\mathrm{pole}(K^- \ell^+ \nu) = 1.81\pm0.03\pm0.02$~GeV/$c^2$
and $m_\mathrm{pole}(\pi^- \ell^+ \nu) = 1.95\pm0.07\pm0.03$~GeV/$c^2$.
Using the measured values of the branching fraction ratio
and the modified pole model with fitted parameter $\alpha_p$, we obtain
\begin{equation}
f_+({D^0 \to \pi^- \ell^+ \nu})^2/f_+({D^0 \to K^- \ell^+ \nu})^2\cdot|V_{cd}|^2/|V_{cs}|^2|_{q^2=0} = 0.041\pm 0.003_{\mbox{stat}}\pm 0.004_{\mbox{syst}}.
\end{equation}
in agreement with recent measurements \cite{ref:cleoff,ref:focus2}.
The form factors $f(q^2)$, obtained using the measured $d\Gamma/dq^2$
distribution and the world average of the Br($D^0 \to K \ell \nu$), are in nice
agreement with the predictions of the recent unquenched lattice QCD
results \cite{ref:unquenched}.

\section*{Acknowledgment}
We very much appreciate the suggestions of the
theoreticians T. Onogi and S. Hashimoto who provided the original
motivation for this work.
We thank the KEKB group for the excellent operation of the
accelerator, the KEK cryogenics group for the efficient
operation of the solenoid, and the KEK computer group and
the National Institute of Informatics for valuable computing
and Super-SINET network support. We acknowledge support from
the Ministry of Education, Culture, Sports, Science, and
Technology of Japan and the Japan Society for the Promotion
of Science; the Australian Research Council and the
Australian Department of Education, Science and Training;
the National Science Foundation of China under contract
No.~10175071; the Department of Science and Technology of
India; the BK21 program of the Ministry of Education of
Korea and the CHEP SRC program of the Korea Science and
Engineering Foundation; the Polish State Committee for
Scientific Research under contract No.~2P03B 01324; the
Ministry of Science and Technology of the Russian
Federation; the Ministry of Higher Education, Science and Technology of the Republic of Slovenia;  the Swiss National Science Foundation; the National Science Council and
the Ministry of Education of Taiwan; and the U.S.\
Department of Energy.

\end{document}